\documentclass[]{aastex701}
\usepackage{amsmath}
\usepackage{hyperref}

\begin{document}

\title{The James Webb Space Telescope Absolute Flux Calibration. V. Near-Infrared Camera Wide Field Slitless Spectroscopy}

\author[0000-0003-3382-5941]{Nor Pirzkal}
\affiliation{ESA/AURA}
\email[show]{pirzkal@stsci.edu}  
\author[orcid=0000-0003-4850-9589,sname='Martha Boyer']{Martha Boyer}
\affiliation{Space Telescope Science Institute}
\email[show]{mboyer@stsci.edu}  
\author[orcid=0000-0003-0894-1588,sname='Russel E. Ryan']{Russel E. Ryan Jr.}
\affiliation{Space Telescope Science Institute}
\email[show]{rryan@stsci.edu}  

\collaboration{all}{The JWST NIRCam Team}

\begin{abstract}
We present the absolute flux and wavelength calibration of the \textit{James Webb Space Telescope} (\textit{JWST}) Near-Infrared Camera (NIRCam) Wide Field Slitless Spectroscopy (WFSS) mode. Each of NIRCam’s two modules (A and B) provides independent long-wavelength (LW) grism spectroscopy over the $2.4$–$5.0~\mu$m range, with orthogonally oriented R and C grisms. Using commissioning and calibration data from programs 
01076, 01536, 01537, 01538, 01479, 01480, 04449, 04498, 06606, and 06628, we have measured the field-dependent geometry and wavelength dispersion of both first- and second-order spectra across the full detector area. The trace geometry was modeled using two-dimensional third-order polynomials that reproduce the observed spectral positions with an RMS accuracy better than $0.1$~pixel. Wavelength calibration, derived from observations of the planetary nebula SMP LMC 58, achieves a precision of $0.65$–$0.91$~\AA\ for the +1 orders and $0.5$~\AA\ for the +2 orders. Absolute flux calibration, established from observations of the G-type star standard P330E, provides a consistent sensitivity function across all grisms and modules with an absolute flux accuracy of 3\%. The resulting calibration framework defines the geometric, wavelength, and photometric reference for all NIRCam WFSS observations and ensures cross-consistency between modules and grism orientations. These calibrations form the basis for accurate slitless spectroscopy with NIRCam and will support ongoing improvements to the \textit{JWST} calibration pipeline and data products.
\end{abstract}

\section{The NIRCAM WFSS mode} 
NIRCam consists of two identical but completely independent modules, designated A and B, which share only a common backplate. The two modules are mirror images of each other, and each has its own optical path, electronics, and detector system. Within each module, a beam splitter divides the incoming light at approximately $2.4,\mu\mathrm{m}$. Light below this wavelength (the Short-Wave, or SW, channel) is directed to an array of four $2048\times2048$ HgCdTe detectors, while the longer-wavelength light (the Long-Wave, or LW, channel) is sent to a single $2048\times2048$ HgCdTe detector.
When NIRCam operates in the Wide Field Slitless Spectroscopy (WFSS) mode, a grism is inserted into the LW optical path, producing spectra that are projected onto the single LW detector of the active module. Modules A and B observe different regions of the sky, separated by approximately $44''$, and function as independent instruments. Consequently, spectra from Module A do not overlap or affect those from Module B, and vice versa.
Although the SW channels include a dispersive mode using the Dispersed Hartmann Sensor (DHS), this mode is independent of the LW WFSS configuration and is beyond the scope of this paper. Nevertheless, many of the LW WFSS calibration procedures are also relevant for DHS calibration. In typical operations, when the LW channel is used for spectroscopy, the SW channel simultaneously images the same field at wavelengths shorter than $2.5,\mu\mathrm{m}$.
Each NIRCam module provides two LW grisms, oriented orthogonally. The R grisms disperse light horizontally on the LW detector, while the C grisms disperse light vertically. The dispersion directions are inverted between Modules A and B: both R grisms disperse light toward the other module, while both C grisms disperse light downward on their respective LW detectors.
In WFSS mode, every source within the effective field of view of a NIRCam module produces a dispersed spectrum. Multiple sources therefore yield overlapping spectra, substantially increasing the complexity of data analysis. The LW grisms were designed with a nominal dispersion of $10,\text{\AA}$ per pixel. Combined with the full LW bandpass of approximately $2.4$–$5.0,\mu\mathrm{m}$, this produces spectra over $2500$ pixels long—exceeding the detector’s field of view. To limit spectral overlap, broad or medium-band filters can be paired with any NIRCam grism to restrict the effective wavelength coverage, thereby shortening the spectral traces and reducing confusion from overlapping spectra.

\section{Data and initial processing}
Following the successful launch of \textit{JWST}, WFSS observations were initially obtained during the observatory's commissioning phase and have subsequently been acquired in each observing cycle as part of dedicated calibration programs. The data analyzed in this paper are drawn from proposals PIDs~01076, 01479, 01480, 04358, 04449, 06628. These data were 
processed using the JWST pipeline, version 1.19.0. All imaging data were processing through the pipeline Stage 1, Stage 2, and Stage 3 to create imaging mosaics (i2d) of each observations. The WFSS observations were processed using Step 1 with the addition of the assign\_wcs() step which populates a 
standard header description of the world coordinate system for convenient access by standard code. We also applied the broad-band flat-field of the cross filter used to acquire the WFSS observations using the pipeline flatfield\_step().

\section{Calibration}
The calibration of the NIRCam WFSS mode broadly follows the procedures established for the calibration of the WFSS modes on the \textit{HST} instruments ACS \citep{Koekemoer2007} and WFC3 \citep{McLaurin2009}. The process involves several independent calibration steps, beginning with the characterization of the geometry of the spectral traces. This geometry is described by a polynomial model that defines the shape of the dispersed spectra produced by sources within the field of view. Specifically, for a source located at $(x_0, y_0)$, the model determines where light of a given wavelength $\lambda$ falls on the detector relative to the undispersed position $(x_0, y_0)$ of the source. 

The coordinates $(x_0, y_0)$ represent the location where the source \emph{would have been measured} if the pupil wheel contained a filter rather than a grism. As discussed in \citet{Pirzkal2017}, this formalism can be generalized so that the relationship between the source position and the location of light on the detector is expressed by three equations (and their inverses):

\begin{subequations}\label{forward}
\begin{align}
\delta_x = x' - x_{0} = f_x(x_{0},y_{0};t)\\
\delta_y = y' - y_{0} = f_y(x_{0},y_{0};t)\\
\lambda = f_\lambda(x_{0},y_{0};t)
\end{align}
\end{subequations}

where $t$ is a free parameter that can be chosen appropriately. In cases where the spectral traces are nearly linear, $t$ can be related to the wavelength $\lambda$, and $f_\lambda$ is chosen such that $0 < t < 1$ over the wavelength range covered by the dispersed spectra.

While $f_x$, $f_y$, and $f_\lambda$ can, in principle, take any functional form, they are allowed to depend on both the source position within the field, $(x_0, y_0)$, and on $t$ or $\lambda$.

In the following sections, we describe in detail the procedure followed to perform the latest recalibration of the NIRCam WFSS mode. This work extends and improves upon previous calibrations, which were performed using the same methodology but with less \emph{a priori} knowledge of the trace geometry and wavelength dispersion of the NIRCam grisms. Initial pre-launch estimates assumed near-linear dispersion and a constant wavelength dispersion of approximately $10$~\AA\ per pixel. 

Following the successful launch of \textit{JWST}, these estimates were refined using observations obtained under PID~01076, a program designed to measure the relative positions of spectral traces using a few stellar targets, determine the wavelength dispersion from observations of the planetary nebula SMP LMC 58 in the  Large Magellanic Cloud (IRAS 05248-7007 05:24:20.75 -70:05:01.60), and calibrate flux using the G-type star P330E (GSC 02581-02323, 16:31:33.81 +30:08:46.40). These observations provided limited coverage over the two NIRCam module fields of view, yet allowed for the measurement of a field-dependent variation in both the trace geometry and wavelength dispersion. 

This initial calibration achieved an accuracy of approximately $0.2$~pixel in predicting the exact location of spectral traces given knowledge of the source positions in the field, and a wavelength calibration accuracy of approximately $2.5$~\AA. However, the limited dataset left some regions of the field, such as the lower corners of the detectors, significantly less well calibrated, particularly in some corners of the field-of-view. In the following paper, we describe in some details the methodology we followed to improve the NIRCAM WFSS calibration. What is described here was actually performed iteratively over a dozen times, as our understanding of the WFSS mode improved. 

\section{Trace Geometry}\label{sec:geometry}
\subsection{Parametrization}
In many modern instruments, including NIRCam, the spectral traces are specifically designed to be approximately linear, but their shapes and relative position to sources vary across the field of view. In this work, we refer to this variation as the ``field dependence'' of the grism disperser and associated optics. 

Calibrating the trace geometry involves determining the simplest function that accurately describes the shape of the spectral traces. This is commonly achieved using a two-dimensional polynomial $P_{n,m}(x,y,t)$ of order $(n,m)$, where the $n^{\rm th}$-order term describes the shape of the trace along the detector (e.g., $n=1$ corresponds to linear traces, $n=2$ to quadratic, and so on), and the polynomial coefficients themselves vary with the source position $(x_0, y_0)$ across the detector. This position dependence is typically modeled using an $m^{\rm th}$-order 2D polynomial, such that:

\begin{subequations}
\begin{align}
P_{n,m}(x,y,t) = \sum_{l=0}^{n} t^l \times \left(\sum_{k=0}^{m} a_{n,m} \times x^i y^j\right)  & \ \ \textrm{where}\ i + j \leq k
\end{align}
\end{subequations}
so that for example a $2^{nd}$\ order trace (quadratic trace) with a $2^{nd}$\ order 2D polynomial field dependence is:
\begin{subequations}
\begin{align}
\begin{split}
P_{2,2}(x,y,t) &= a_{0,0} + a_{0,1} \times x + a_{0,2} \times t + a_{0,3} \times x^2 + a_{0,4} \times  x \times y  +  a_{0,5} \times y^2
\\&
+ t \times (a_{1,0} + a_{1,1} \times x + a_{1,2} \times t + a_{1,3} \times x^2 + a_{1,4} \times  x \times y  +  a_{1,5} \times y^2 ) 
\\&
+ t^2 \times (a_{2,0} + a_{2,1} \times x + a_{2,2} \times t + a_{2,3} \times x^2 + a_{2,4} \times  x \times y  +  a_{2,5} \times y^2)
\end{split}
\end{align}
\end{subequations}

\subsection{Data}

Measuring spectral traces can, in principle, be performed using any source in the field; however, both locating a source and fitting its spectral trace are more straightforward when using stars, as these point-like objects produce sharper traces in the cross-dispersion direction. To maximize the number of sources used for measuring spectral traces, we focused on spectra that are likely to be minimally contaminated by neighboring fainter or brighter sources. Using the then-current knowledge of the dispersion relation, we quantitatively estimated the contamination at each pixel of each spectrum. Starting with a manually determined relation (as described above) allowed for an initial selection of sources, increasing the number of usable measurements. This iterative process—where improved calibration informed the selection of additional uncontaminated traces—was repeated several times, progressively increasing the field coverage and overall accuracy of the trace geometry model.

We used data from proposals 01076, 01480, 01479, 04449, and 06628. These programs yielded a total of 112 and 96 observations for Module~A using grisms R and C, respectively, and 96 and 80 observations for Module~B using grisms R and C, respectively, resulting in a total dataset of 384 observations. Except for a few observations using the F356W filter as a cross-filter, the majority of WFSS observations were obtained using the F322W2 or F444W filters, providing shorter spectra with blue and red wavelength coverage, respectively.

Usable spectral traces were identified by first processing the associated direct images and running the JWST pipeline Stage~3 to produce i2d mosaic files. Source Extractor was then used to generate catalogs and segmentation maps for each field. Finally, the Simulation-Based Extraction (SBE) code \citep{Pirzkal2024} was employed to forward-model the direct images, producing fully simulated WFSS images for each of the 384 observations. These simulations were used to compute the fractional contamination at each pixel along each trace, and only portions of spectra with less than 4\% contamination were retained—sufficiently low to avoid significant bias in determining the centroid of traces in the cross-dispersion direction.

Because trace calibration is performed relative to the source position in the field, $(x_0, y_0)$, it is crucial to avoid blended traces, where overlapping spectra could introduce systematic errors by effectively measuring the position of a neighboring spectrum rather than the target. The number of identified traces for each combination of module, grism, and direct filter is summarized in Table~\ref{traces}. With several thousand +1 order spectra distributed across the detectors, the data provided good spatial coverage over the entire LW detectors. We also identified several hundred +2 order spectra, which are visible only when using the bluest cross-filters (F322W2 or bluer). As the +2 order spectra are significantly fainter, they are more susceptible to contamination from both +1 and other +2 order spectra. Examples of an observation and SBE simulation are shown at the end of this paper.

\begin{deluxetable}{ccccc}
\tablecaption{Number of spectra measured for each combination of Module, Grism, and cross filter.\label{traces}}
\tablehead{\colhead{Module} & \colhead{Grism} & \colhead{Filter} & \colhead{+1 Spectra} & \colhead{+2 Spectra} }
\startdata
A & R & F322W2 & 3006 & 317\\
A & R & F444W & 3265 & \\
A & C & F322W2 & 2091 & 303\\
A & C & F444W & 3180 & \\
B & R & F322W2 & 1995 & 177\\
B & R & F444W & 2179 &  \\
B & C & F322W2 & 1496 & 177\\
B & C & F444W & 2213 &\\
\enddata

\end{deluxetable}

\subsection{Spectral Trace Measurements}
\begin{figure*}
\center
\includegraphics[width=6.5in]{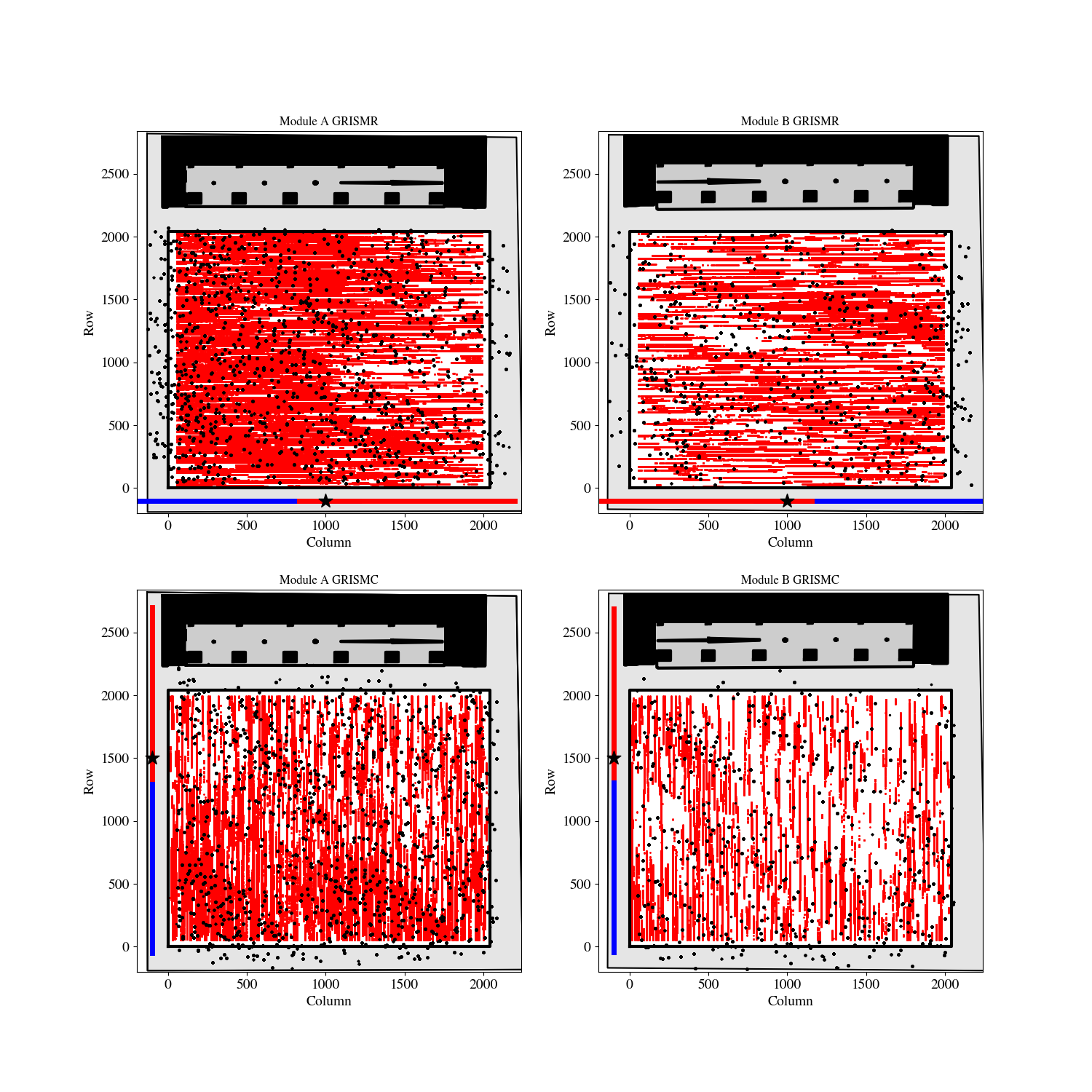}
\caption{This figure shows the source positions (black dots) along with the measured segments of the dispersed +1 order traces that were used to calibrate their shape. The coronagraph footprint above the detector is indicated by a black square, with its semi-transparent subtraction shown in dark gray. The approximate extent of the pick-off mirror (POM), which defines the effective field of view of the WFSS mode, is shown in light gray. Our measurements provide good spatial coverage across both the detector and the effective field of view. The spectral trace segments are shown in red and include data obtained with both F322W2 and F444W filters. For an object located at the position indicated by the large black star, the extents of the F322W2 and F444W traces are highlighted with thick blue and red lines, respectively.
 \label{Trace_FOV}}
\end{figure*}

\begin{figure*}
\center
\includegraphics[width=6.5in]{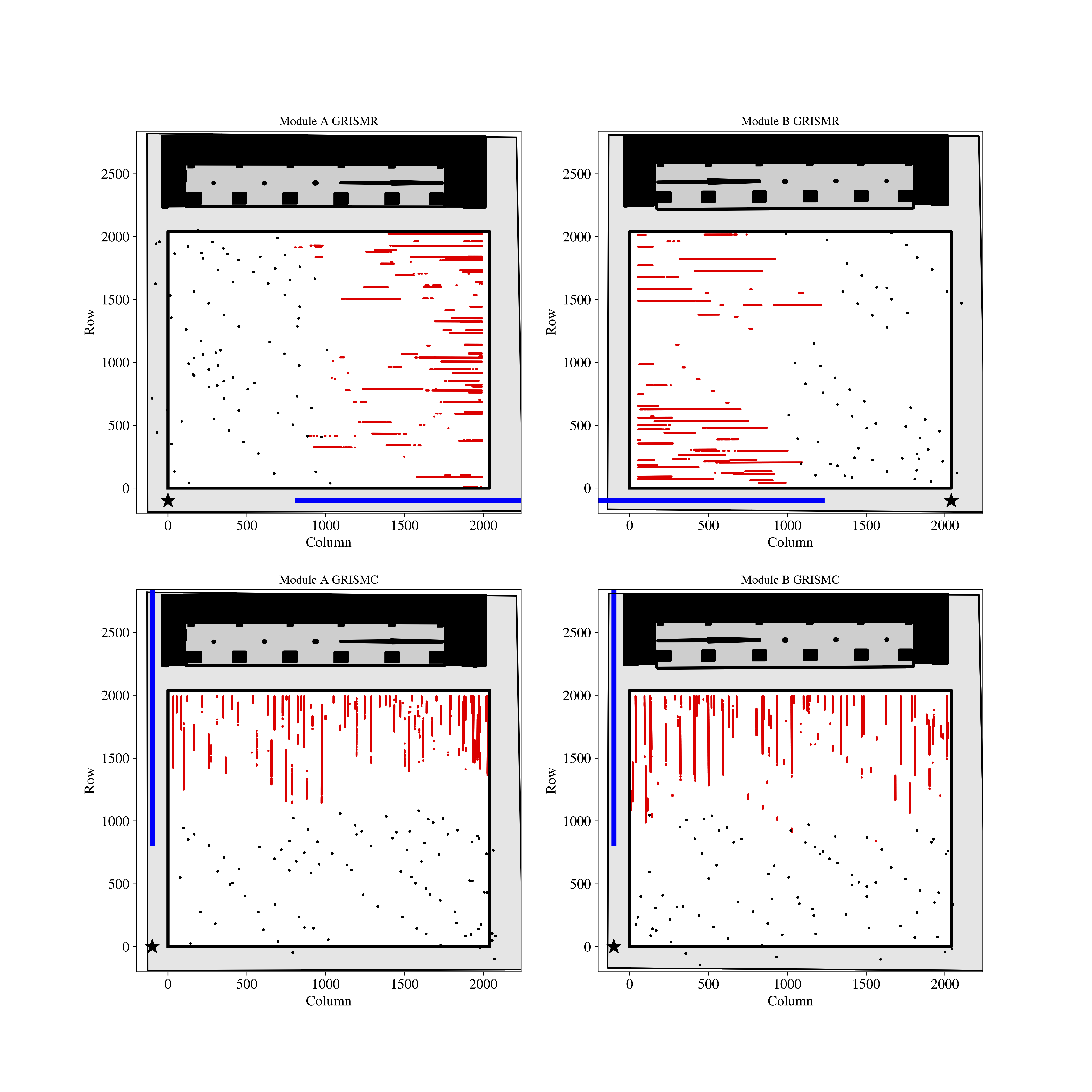}
\caption{This figure is analogous to Figure~\ref{Trace_FOV}, but shows the +2 order spectra. Only F322W2 data are included, as the +2 order of the F444W spectra falls outside the detector.\label{Trace_FOV2}}

\end{figure*}

The positions of the spectral traces were determined by fitting the light distribution in the cross-dispersion direction. At each integer position along the trace, the median profile within a $\pm 5$~pixel window was modeled using a Gaussian function plus a continuum component. The centroid of the fitted Gaussian, $(\delta x, \delta y)$, relative to the known source position $(x_0, y_0)$, was recorded at each step. The resulting measured dispersed traces are shown in Figures~\ref{Trace_FOV} and \ref{Trace_FOV2}.

\subsection{Fitting}
\begin{figure*}
\center
\includegraphics[width=4.5in]{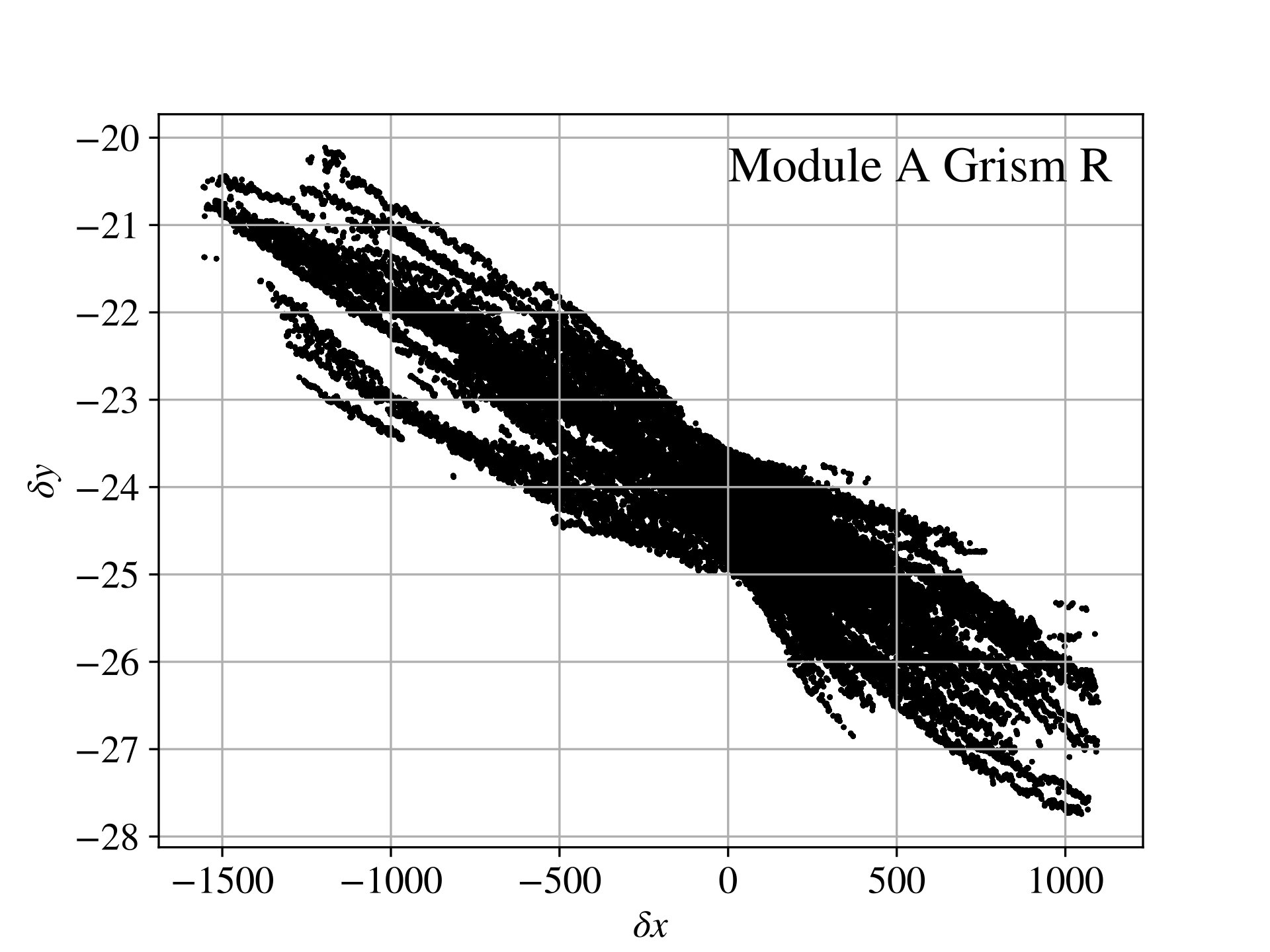}
\caption{All traces shown in Figure~\ref{Trace_FOV} for Module~A and Grism~R, plotted relative to the source position $(x_0, y_0)$. The dispersed traces exhibit a wide range of shapes and curvature. \label{Trace_example}}

\end{figure*}

\begin{figure*}
\center
\includegraphics[width=6.5in]{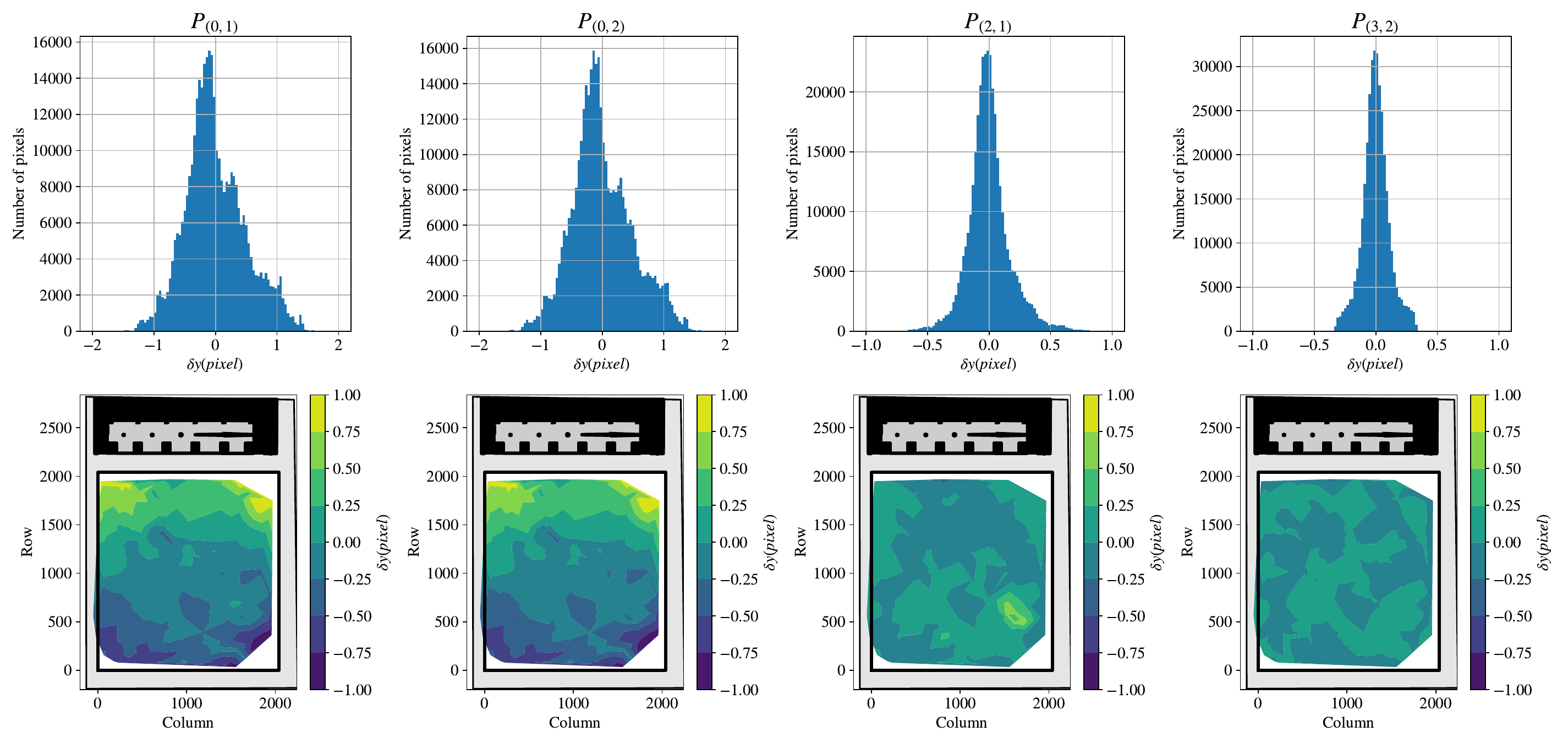}
\caption{Impact of the polynomial order on the residuals of the trace geometry fit using increasingly higher-order $P_{n,m}(x,y,t)$ polynomials. As shown, using a simple first-order trace with no field dependence ($P_{0,1}(x,y,t)$) results in large residuals of several pixels, as the variations in trace geometry are not captured. A second-order trace without field dependence ($P_{0,2}(x,y,t)$; second panel) also fails to model the geometry accurately. Only by including field dependence (third and fourth panels from the left) are the residuals significantly reduced to a small fraction of a pixel, uniformly across the field of view. \label{ModAR_Ps}}

\end{figure*}

Figure~\ref{Trace_example} shows all spectra for Module~A with Grism~R obtained using F322W2 and F444W as cross-filters, plotted as the displacement in the cross-dispersion direction, $\delta y$, versus the displacement in the dispersion direction, $\delta x$, measured relative to the source position $(x_0, y_0)$. As illustrated in this figure, the offsets and curvature of the spectral traces vary significantly across the field of view. To model this distribution of dispersed spectra, we employed the polynomial $P_{n,m}(x,y,t)$, varying $n$ (the 2D field-dependence order) and $m$ (the order of the individual trace). Figure~\ref{ModAR_Ps} presents both histogram and contour plots of the residuals for different choices of $P_{n,m}(x,y,t)$. As shown, good fits were obtained using $P_{3,2}(x,y,t)$, which models parabolic traces with a third-order 2D field dependence.

Using the relatively uncontaminated traces identified previously, we iteratively fitted all measurements $(x_0, y_0, \delta x, \delta y)$ to the polynomial $P_{3,2}(x,y,t)$. This polynomial order was chosen as it minimized the RMS of the fit without overfitting the data. Measurements for each combination of Module, filter, and grism were fitted separately, starting with several hundred thousand trace measurements. A two-stage iterative scheme was employed to remove outliers. First, a fit was performed, residuals were computed, and $5\sigma$ outliers were removed. This step was repeated iteratively until no further outliers remained. The procedure was then repeated multiple times while progressively lowering the rejection threshold in steps of $-0.25\sigma$ down to $2\sigma$. Approximately 10\% of measurements were rejected using this method, while maintaining uniform spatial coverage across each Module.

Initially, trace fitting was performed separately for F322W2 and F444W data. However, the resulting solutions agreed to within $\le 0.1$~pixel near the detector center. Therefore, we combined the F322W2 and F444W traces to produce a single solution that is independent of the cross-filter used. The same methodology was applied to both +1 and +2 order data.

\subsection{Accuracy of Polynomial Model}
\begin{figure*}
\center
\includegraphics[width=6.5in]{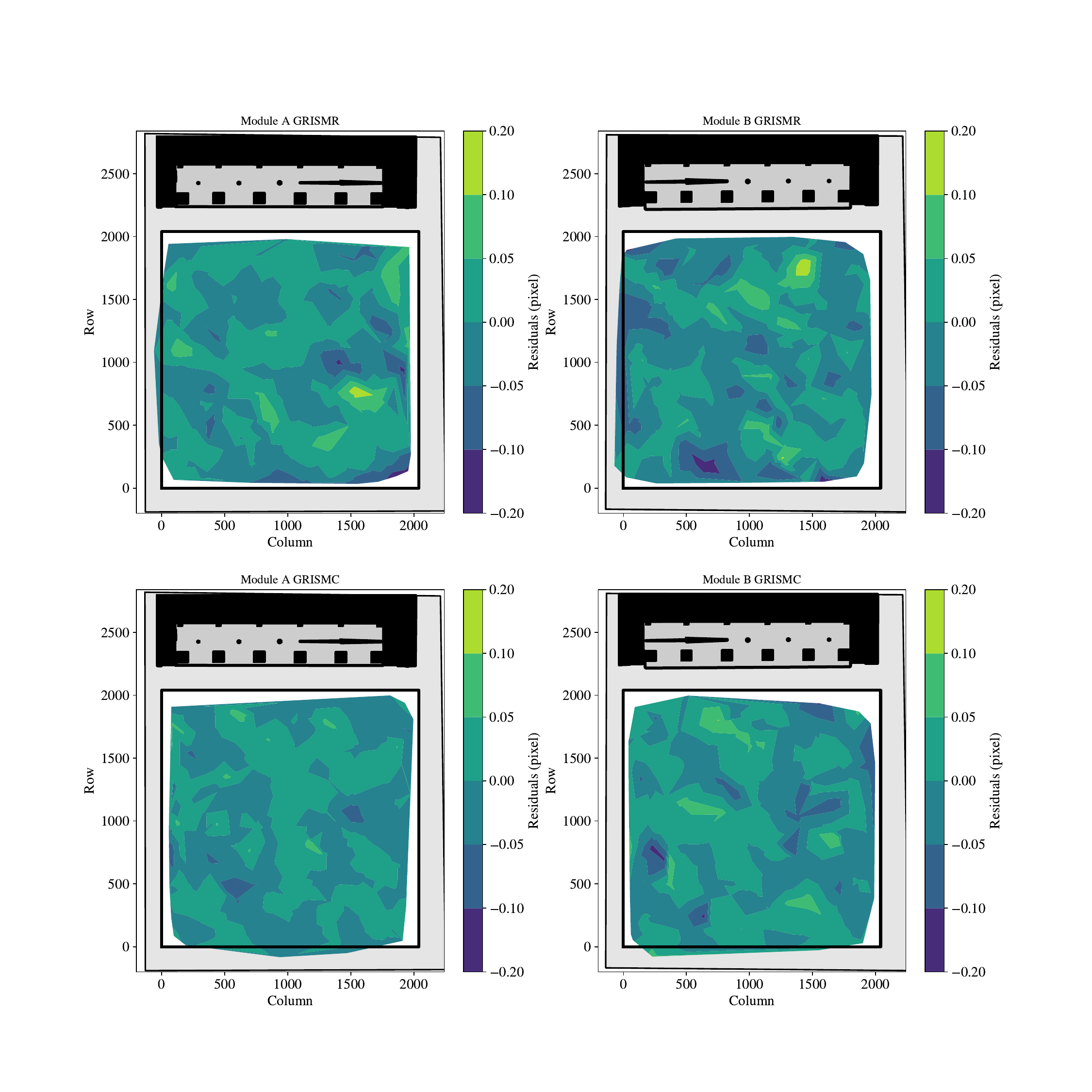}
\caption{Contour plots of the mean residuals between our fitted model and the observations for the +1 order traces. As shown, there is little to no field dependence in the RMS of the fit, the spatial coverage is uniform, and the mean error remains within a small fraction of a pixel across the entire detector. \label{Trace_res_FOV}}

\end{figure*}

\begin{figure*}
\center
\includegraphics[width=6.5in]{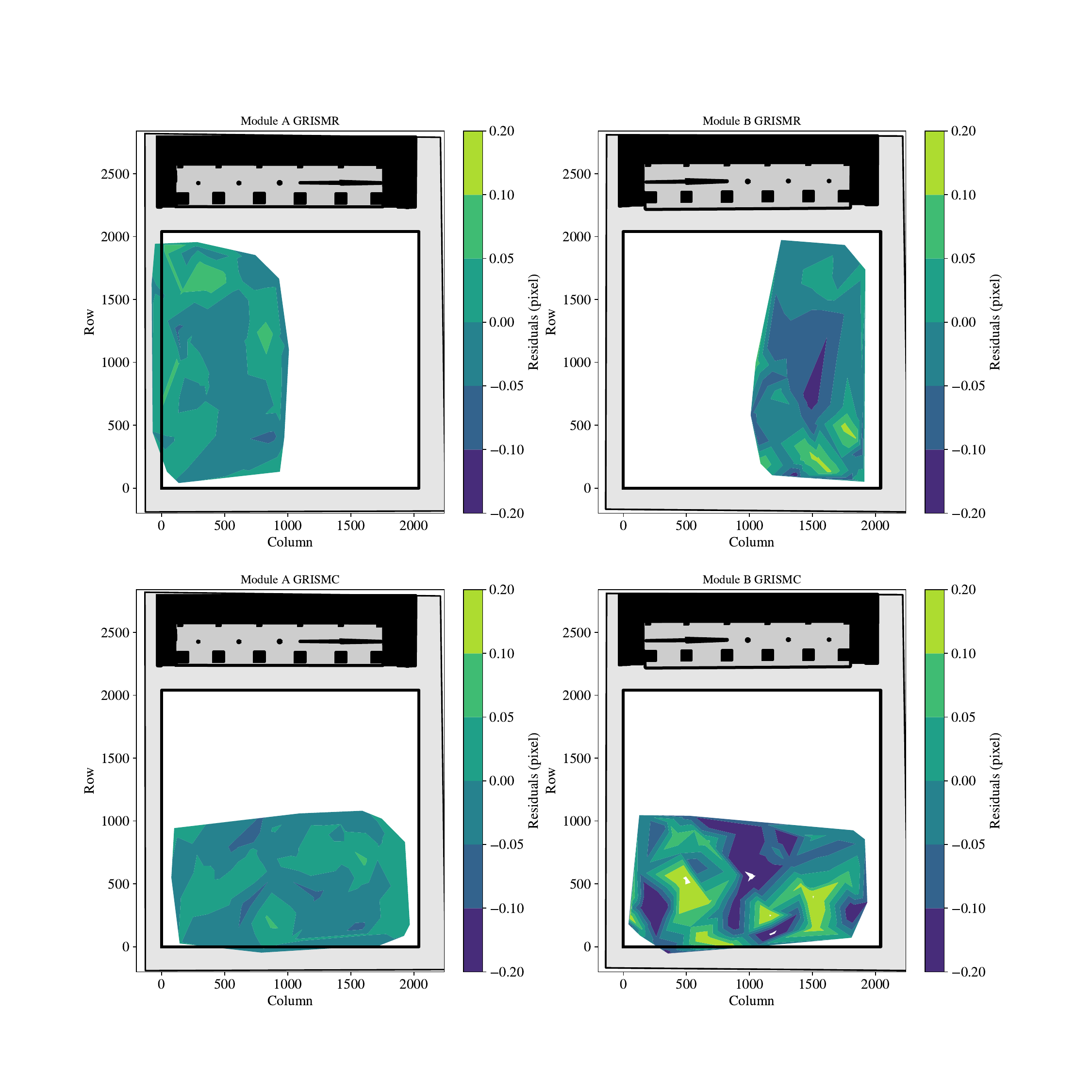}
\caption{Similar to Figure~\ref{Trace_res_FOV}, but for the +2 order traces. Although the available data are significantly fewer than for the +1 order, our model still fits the measurements to within a fraction of a pixel. \label{Trace_res_FOV2}}
\end{figure*}

As shown in Figure~\ref{Trace_FOV}, we were able to measure and fit the +1 order traces for all grism–module combinations across a large number of positions, fully covering the field of view. Comparisons between the measurements and predictions from our fitted model are presented in Figures~\ref{Trace_res_FOV} and \ref{Trace_res_FOV2} for the +1 and +2 orders, respectively. No field dependence is observed in the residuals, indicating that the 2D polynomial fit adequately describes the trace geometry. The mean residuals across the field are on the order of a few thousandths of a pixel, showing no significant bias or offset between the model and observations. The standard deviation of the residuals is approximately 0.06~pixel. Consequently, the trace models presented here can be expected to predict the location of a dispersed spectrum, given the known position of the source, to better than 0.1~pixel for any combination of module and grism in NIRCam.

\section{Wavelength Calibration}\label{sec:wav}
Following the calibration of the dispersed spectral traces described above, we used observations of a wavelength calibrator with strong emission lines to perform the wavelength calibration of the NIRCam dispersed spectra. The selected source was SMP LMC 58, a planetary nebula in the Large Magellanic Cloud (LMC). This object exhibits several strong emission lines between 2.5 and 5~$\mu$m, enabling determination of the spectral dispersion at multiple points along the dispersed traces and allowing us to characterize any nonlinearity in the wavelength solution. As for the trace geometry calibration, we allowed for a 2D field dependence in the wavelength solution. 

\begin{figure*}
\center
\includegraphics[width=6.5in]{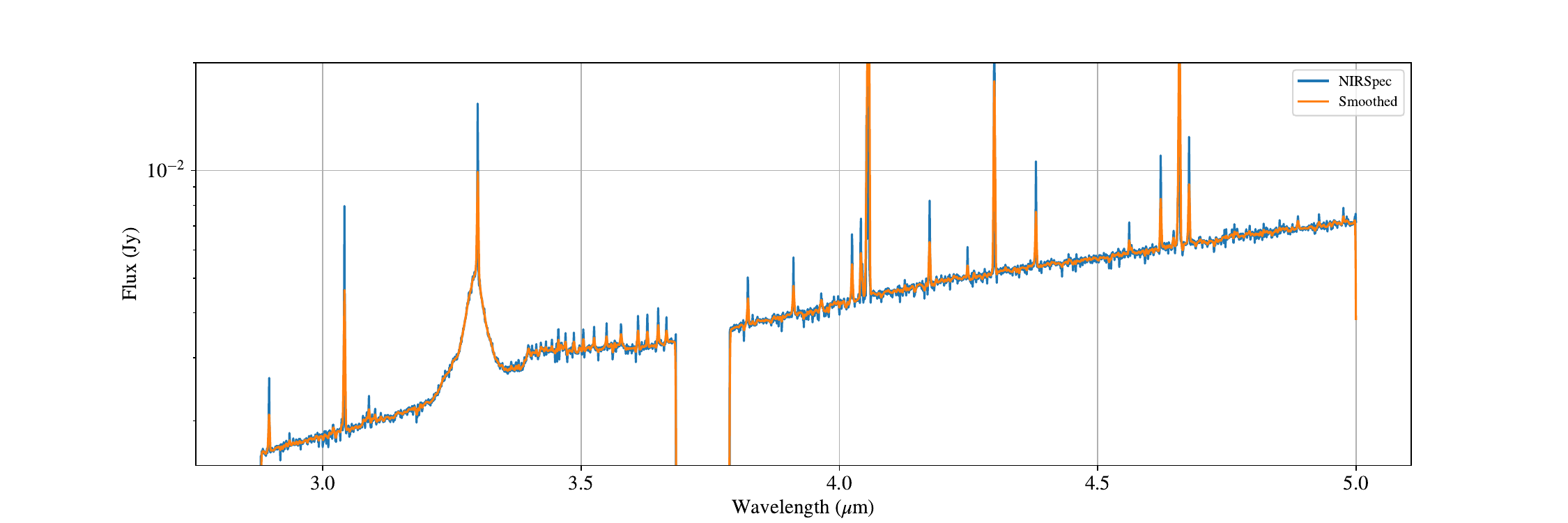}
\caption{Reference spectra of our wavelength calibrator, SMP LMC 58.\label{refwave}}
\end{figure*}

The wavelength calibrator was observed at different positions on the detectors of both Module~A and Module~B using the R and C grisms. To establish the fiducial wavelengths of emission lines in the spectra of our target, we used a set of high-resolution NIRSpec spectra from program PID~01125 as reference \citep{Boker23}. These spectra were obtained using the F170LP/G235H and F290LP/G395H filters with high-resolution gratings (R~$\approx 2700$). The NIRSpec spectra provide accurate emission-line ratios and account for the heliocentric velocity difference between the LMC and the Milky Way. These reference spectra are shown in Figure~\ref{refwave}.

To match the resolution of the NIRCam WFSS spectra (R~$\approx 1600$, $10$~\AA/pixel), the NIRSpec spectra were smoothed using a Gaussian convolution. We then identified a set of emission lines in the reference spectra and used the same lines in our WFSS observations. The identified lines are listed in Table~\ref{tab:refwave}.

\begin{deluxetable}{c}
\tablecaption{List of emission lines used for the wavelength calibration of NIRCam WFSS observations, as measured from the spectra shown in Figure~\ref{refwave}.
\label{tab:refwave}}
\tablehead{\colhead{Fiducial Observed Wavelengths ($\mu m)$}}
\startdata
2.628188 \\
2.760727 \\
2.875560 \\
3.041741 \\
3.299745 \\
4.041627 \\
4.055815 \\
4.174464 \\
4.299690 \\
4.380250 \\
4.621885 \\
4.657861 \\
4.676662 \\
\enddata

\end{deluxetable}

\subsection{Data and Parametrization}
The wavelength calibration employs the same type of polynomials used for the trace geometry. The number of observations available for wavelength calibration of the NIRCam WFSS mode is significantly smaller, as it relies on a single calibrator that must be placed at different positions on the detectors of Modules~A and~B. Since the commissioning of \textit{JWST} and NIRCam, several calibration programs have observed SMP LMC 58 a total of 240 times (Programs~01076, 01479, 01480, 04449, 06628) using F322W2 and F444W as cross-filters.

\subsection{Measurements}\label{sec:wave}
The wavelength calibration was performed using the trace geometry described above. One-dimensional spectra were extracted by first producing 2D subimages containing the target spectrum with a cross-dispersion width of 8 pixels. Depending on the grism orientation (R or C), these 2D subimages were summed along the cross-dispersion direction. Emission lines were then identified in the resulting 1D spectra and fitted with a Gaussian profile plus a local linear continuum. The measured positions of the emission lines in the 1D spectra were parameterized relative to the location of the calibrator $(x_0, y_0)$ along the dispersion direction (i.e., $\delta x$ for grism R and $\delta y$ for grism C). Each measurement thus consisted of the detector coordinates $(x_0, y_0)$, the fiducial wavelength $\lambda_0$, and the offset from the source position in the extracted 1D spectra ($\delta x$ or $\delta y$).

The wavelength calibrator, SMP LMC 58, saturated in many of the early F322W2 and F444W observations. This issue was mitigated in later observing cycles by adding the F405N narrow-band filter to F444W imaging exposures. We found excellent astrometric consistency between images obtained using F322W2, F356W, and F444W, as well as with simultaneously obtained SW observations, with agreement at the $\sim 0.1$~pixel level. This implies that imaging obtained with F322W2, F444W, or F444W+F405N can be used to accurately determine the positions of stars in the field. In observations where SMP LMC 58 was saturated, we matched GAIA~DR3 stars near the target to derive an affine transformation, which allowed us to infer the position of the saturated calibrator while accounting for proper motions. This procedure enabled locating the calibrator to better than 0.1~pixel in all cases.

\begin{figure*}
\center
\includegraphics[width=6.5in]{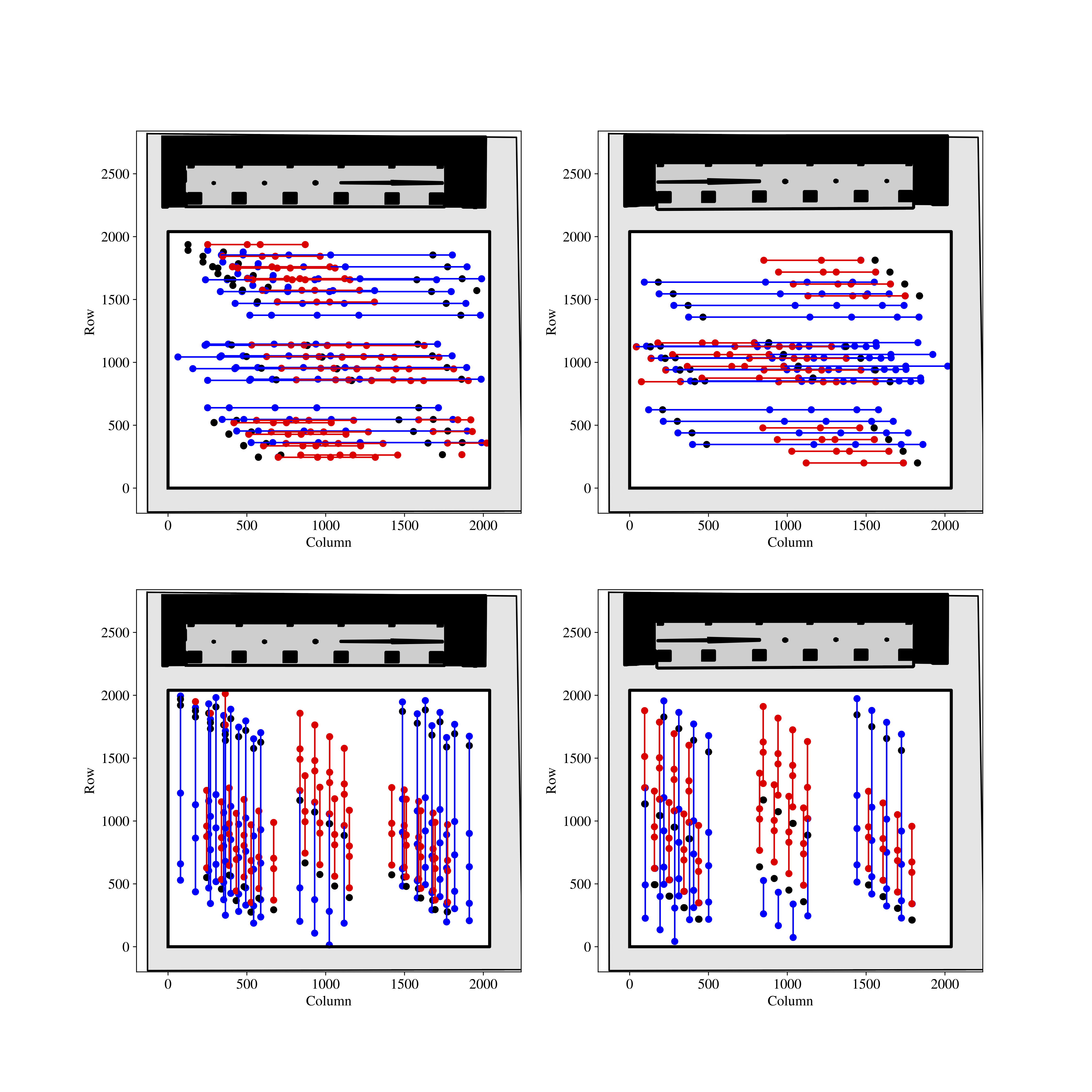}
\caption{Positions of the wavelength calibrator SMP LMC 58 on the LW detectors for the R and C grisms. The object locations are shown with black dots. Approximate spectral traces and the measured emission-line positions are indicated in blue for F322W2 data and red for F444W data. \label{Wave_FOV}}
\end{figure*}

Figure~\ref{Wave_FOV} shows the various positions where SMP LMC 58 was observed on the LW detectors of Modules~A and~B (black dots). The figure also shows the approximate locations of the emission lines measured in the F322W2 spectra (blue dots) and F444W spectra (red dots). Although these measurements are sparser than those used for the trace geometry calibration in Section~\ref{sec:geometry}, they provide adequate coverage for each grism across the entire detectors. This will be further improved with additional observation of SMP LMC 58 as part of the JWST NIRCAM calibration and monitoring programs.

\subsection{Emission Lines Fitting}\label{sec:fitting}
The wavelength fitting was performed using emission lines detected in the F322W2 and F444W data simultaneously, following an approach analogous to that described in Section~\ref{sec:geometry} for modeling the shape of the dispersed traces. Heliocentric corrections to the expected wavelengths of the observed emission lines were included, using the value of VELOSYS, the Barycentric correction to the radial velocity) from each individual dataset.

\begin{figure*}
\center
\includegraphics[width=6.5in]{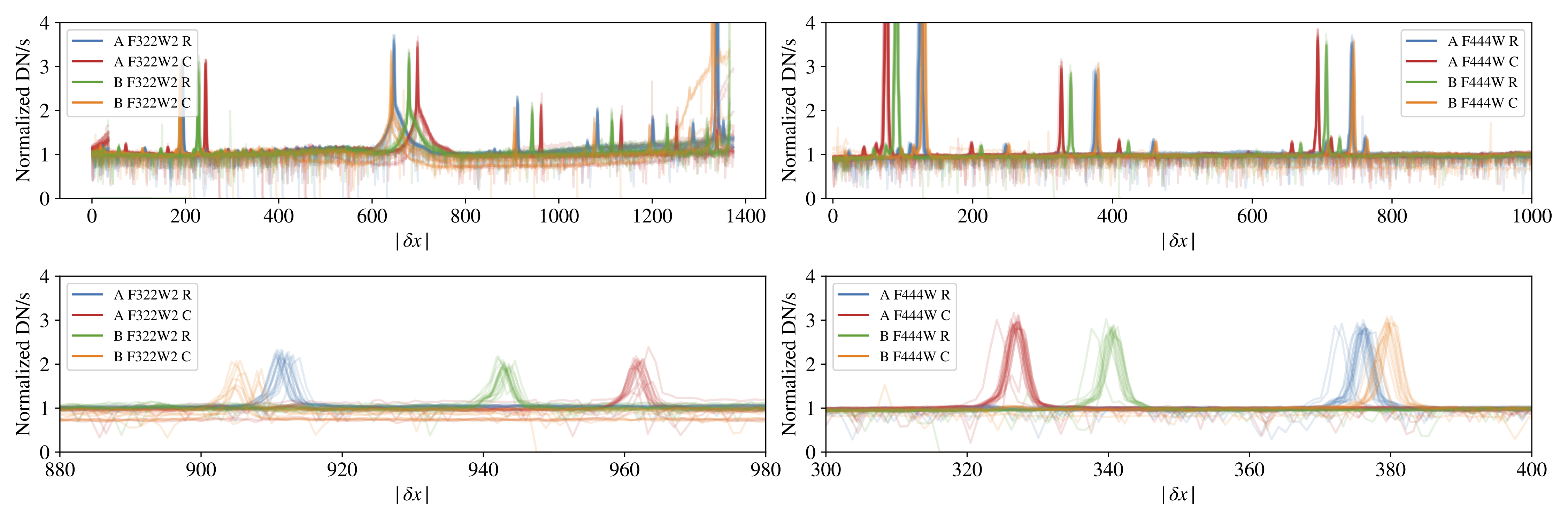}
\caption{Variation in the wavelength dispersion for all combinations of Module (A and B) and grism (R and C). Each combination produces different offsets between the source position $(x_0, y_0)$ and the observed positions of emission lines along the dispersed spectra. The top panels show the full wavelength range of the NIRCam grisms, while the bottom panels display a narrower portion of the spectra. \label{Wave_variation}}
\end{figure*}

Significant differences are observed between spectra extracted from different parts of the field, as well as between the R and C grisms. Figure~\ref{Wave_variation} illustrates how the R (blue) and C (red) spectra of SMP LMC 58 vary as a function of distance from the source position $(x_0, y_0)$. In addition to the systematic differences between the grisms, substantial variation is seen among observations obtained with the same grism and module.

To assess the quality of the wavelength calibration, we consider both the field-dependent variations and the differences between grisms by computing the standard deviation of the derived wavelengths for individual lines. The grism resolution is approximately $10$~\AA\ per pixel. Wavelength calibration involves determining a 2D polynomial $P_{n,m}(x,y,t)$ that accurately models the relationship between wavelength and the parameter $t$ (set initially to $\delta x$ or $\delta y$ depending on the grism).

\begin{figure*}
\center
\includegraphics[width=6.5in]{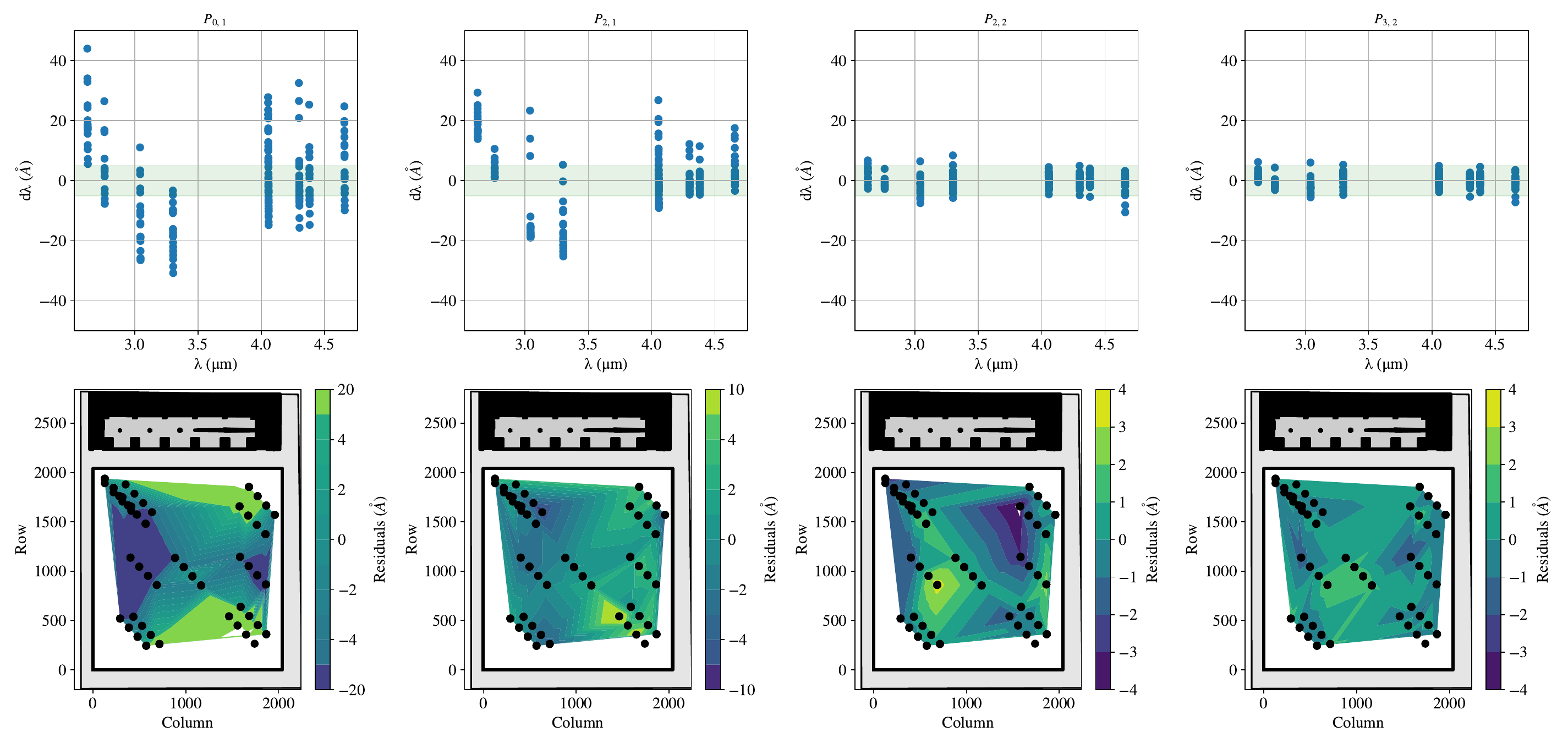}
\caption{Fit residuals for the wavelength solutions using different polynomial models. The left panel shows results for a simple linear relation with no field dependence ($P_{0,1}(x,y,t)$). The two middle panels show fits including 2D field dependence with either a linear or quadratic wavelength relation ($P_{2,1}(x,y,t)$ and $P_{2,2}(x,y,t)$, respectively). The right-most panel shows the accuracy achieved using a third-order field dependence with a quadratic wavelength solution. The nominal dispersion of $10$~\AA/pixel is indicated by the green band in the top plots. \label{Wave_ModAR_Ps}}

\end{figure*}

Figure~\ref{Wave_ModAR_Ps} illustrates the impact of using increasingly higher-order polynomials to model the Module~A, Grism~R observations. As shown, using $P_{2,3}(x,y,t)$ achieves a wavelength calibration accurate to approximately 2~\AA. The final calibration was therefore performed with a $P_{2,3}(x,y,t)$ polynomial, employing a double-iterative procedure similar to that described in Section~\ref{sec:geometry}.

The wavelength dispersion of the NIRCam grisms exhibits significant variation both along the trace and across the field of view. Figure~\ref{Wave_ModAR_Ps} illustrates this by showing the residuals obtained when using a simple polynomial $P_{1,1}(x,y,t)$, which includes no field dependence, compared to a higher-order polynomial $P_{2,3}(x,y,t)$. As the figure demonstrates, variations of several pixels are present with the lower-order model, but these are substantially reduced when using the higher-order polynomial.

\subsection{Accuracy of model}
\begin{figure*}
\center
\includegraphics[width=6.5in]{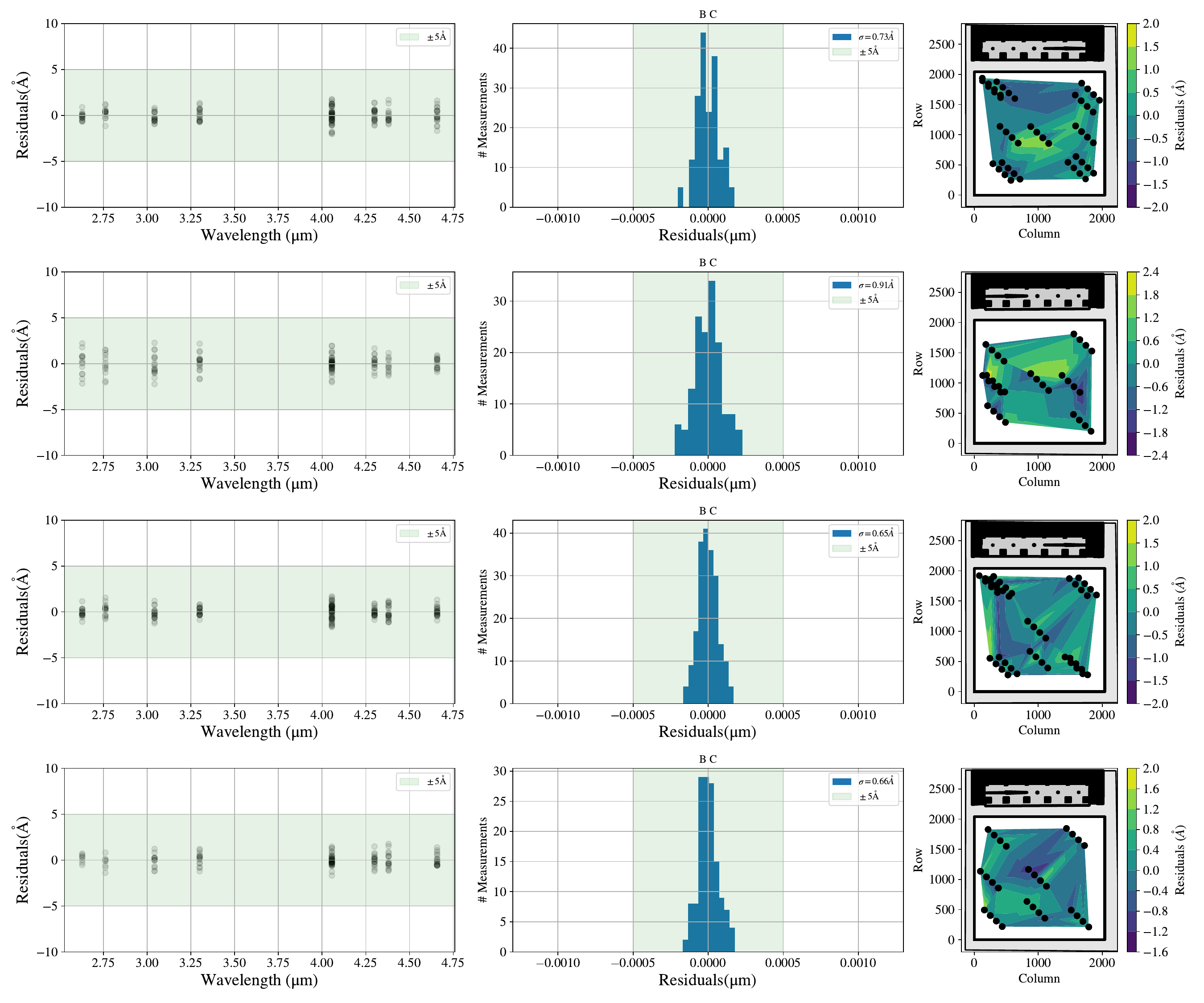}
\caption{Residuals between the fiducial emission wavelengths and the model-predicted wavelengths for the NIRCam +1 order observations of the wavelength calibrator SMP LMC 58. The left panels show the wavelength differences as a function of wavelength, with the nominal pixel dispersion ($10$~\AA) indicated by green bands. The middle panels show histograms of the wavelength residuals. The right panels display the field of view, the spatial coverage of our observations, and a contour plot of the wavelength residuals. The achieved wavelength calibration accuracy is better than 1~\AA. \label{Wave_final}}

\end{figure*}

Figure~\ref{Wave_final} illustrates the accuracy of our final wavelength calibration models by showing the residuals between the fiducial emission line wavelengths and the model-predicted wavelengths for all four combinations of modules and grisms. As the figure demonstrates, full field coverage is achieved (black dots in the rightmost panels), with residuals ranging from 0.65 to 0.91~\AA\ (middle panel) and minimal wavelength dependence (leftmost panel).

\begin{figure*}
\center
\includegraphics[width=6.5in]{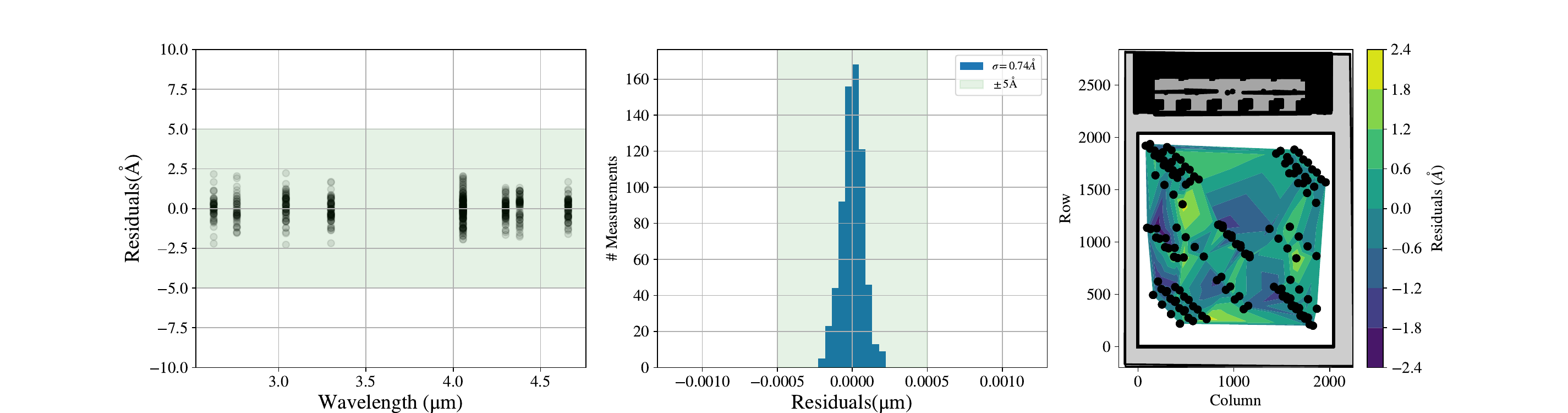}
\caption{Similar to Figure~\ref{Wave_final}, but combining all measurements from all grisms and modules. The calibrated wavelengths show excellent internal consistency, with a typical residual error of $\sigma = 0.74$~\AA. \label{Wave_All}}
\end{figure*}

\begin{figure*}
\center
\includegraphics[width=6.5in]{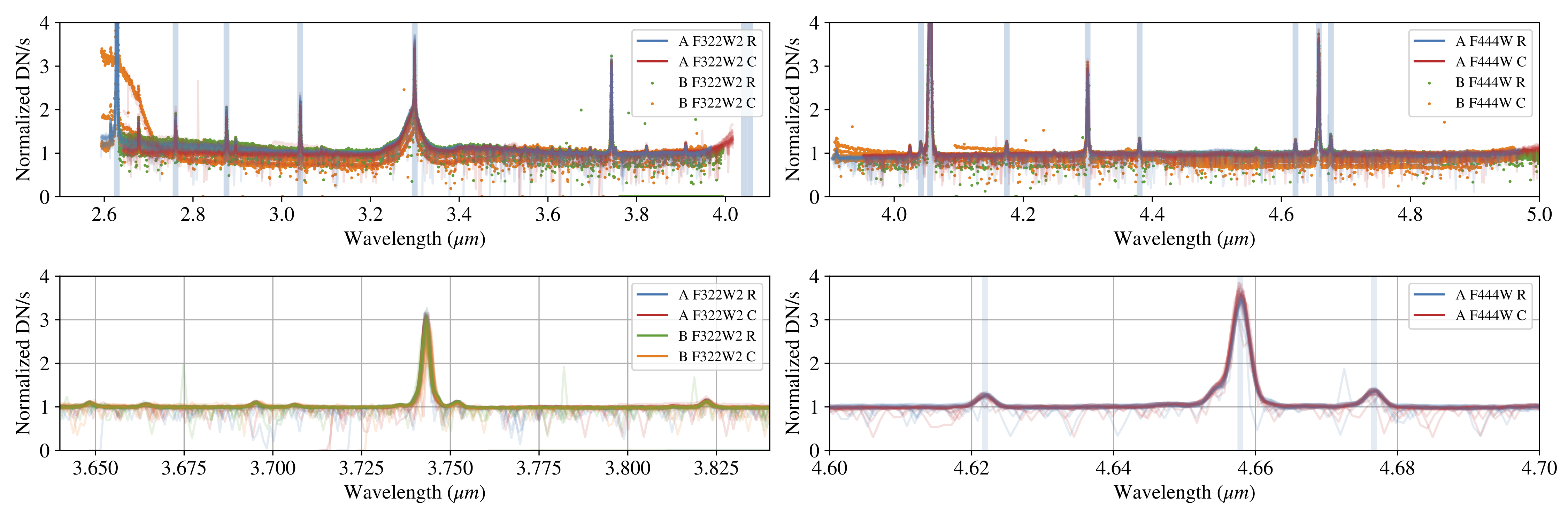}
\caption{The same data shown in Figure~\ref{Wave_variation}, now after applying our wavelength calibration model. After calibration, all extracted spectra are in good agreement. \label{Wave_variation2}}
\end{figure*}

Since all four grisms were calibrated using the same methodology and fiducial line list, we can assess the consistency of the wavelength calibration across all emission lines. Figure~\ref{Wave_All} presents a plot similar to Figure~\ref{Wave_final}, but combining measurements from all grisms and modules. The figure demonstrates that the wavelength solution is consistent across both grisms and modules, with a standard deviation of $\sigma = 0.74$~\AA. Figure~\ref{Wave_variation2} shows the result of applying our wavelength calibration model to the data discussed in Section~\ref{sec:fitting} and shown in Figure~ \ref{Wave_variation}.

\begin{figure*}
\center
\includegraphics[width=6.5in]{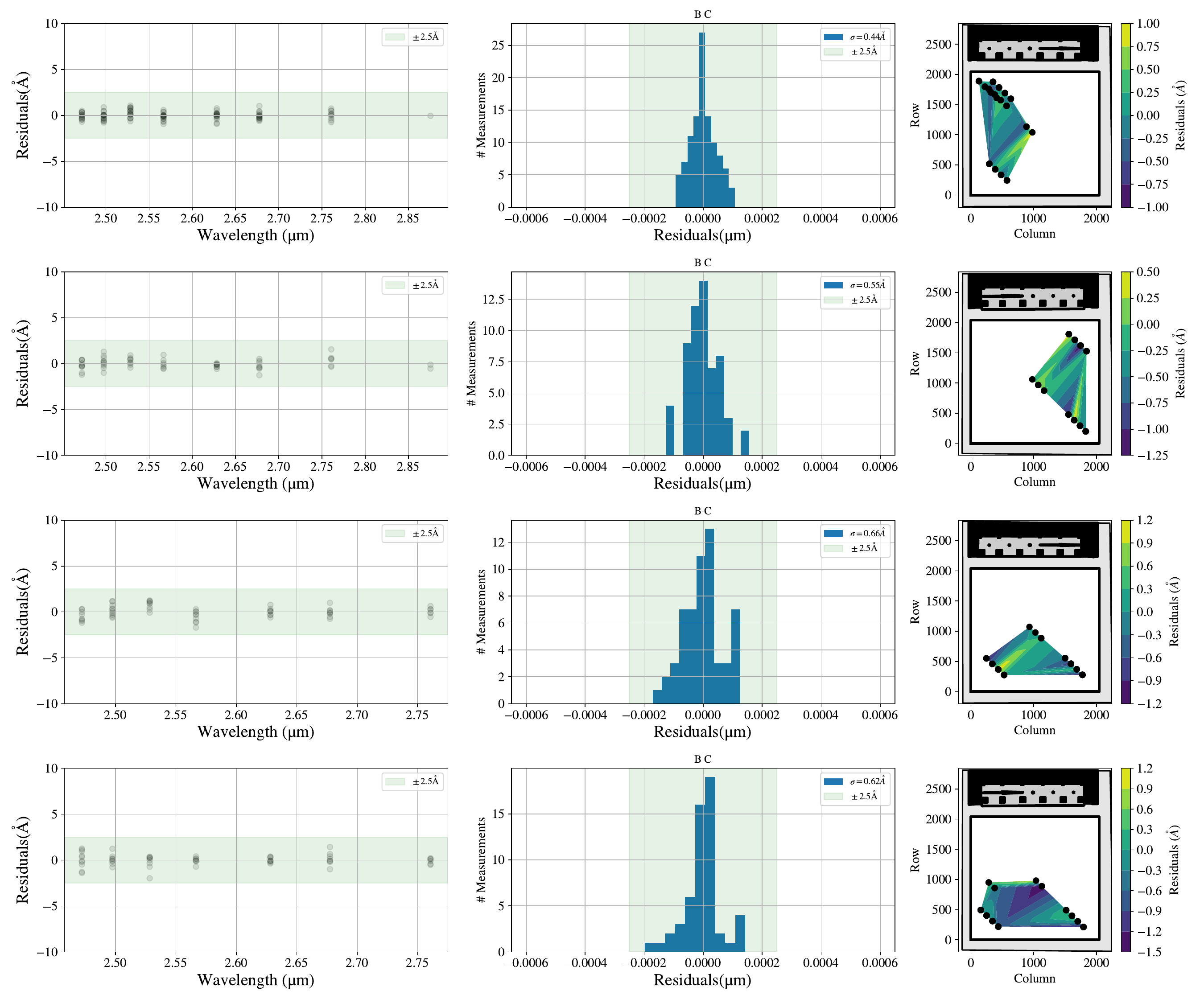}
\caption{Similar to Figure~\ref{Wave_final}, but for the +2 order spectra. The left panels show the differences between model-predicted and fiducial wavelengths as a function of wavelength. The middle panels display histograms of the wavelength residuals. The right panels show the field of view, spatial coverage of the observations, and a contour plot of the residuals. The wavelength calibration error is approximately 0.5~\AA, and the nominal pixel dispersion ($5$~\AA) is indicated by the green bands in the leftmost and middle panels. \label{Wave_final2}}
\end{figure*}

\begin{figure*}
\center
\includegraphics[width=6.5in]{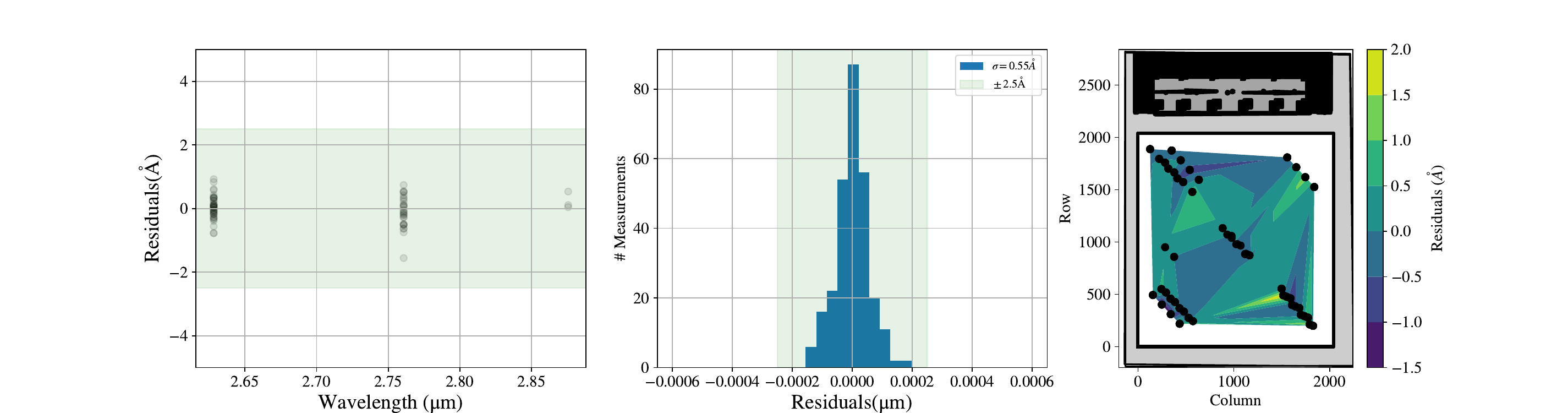}
\caption{This Figure is similar to Figure \ref{Wave_All} but for the +2 orders.\label{Wave_All2}}
\end{figure*}

The same procedures were applied to calibrate the +2 order, with two important caveats: (1) the +2 orders are only present on the detector for wavelengths blueward of $3~\mu$m, and the spectra are otherwise truncated; (2) the number of +2 order emission lines available for calibration is significantly smaller than for the +1 order spectra. Figures~\ref{Wave_final2} and \ref{Wave_All2}, analogous to Figures~\ref{Wave_final} and \ref{Wave_All}, show the residuals of the +2 order wavelength calibration. Although a field-dependent dispersion model was implemented to account for variations across the detector, the +2 orders are only detected for sources located in a small region, so the field dependence is predominantly along the cross-dispersion direction.

\section{Flux Calibration of +1 order}
\begin{figure*}
\center
\includegraphics[width=4.5in]{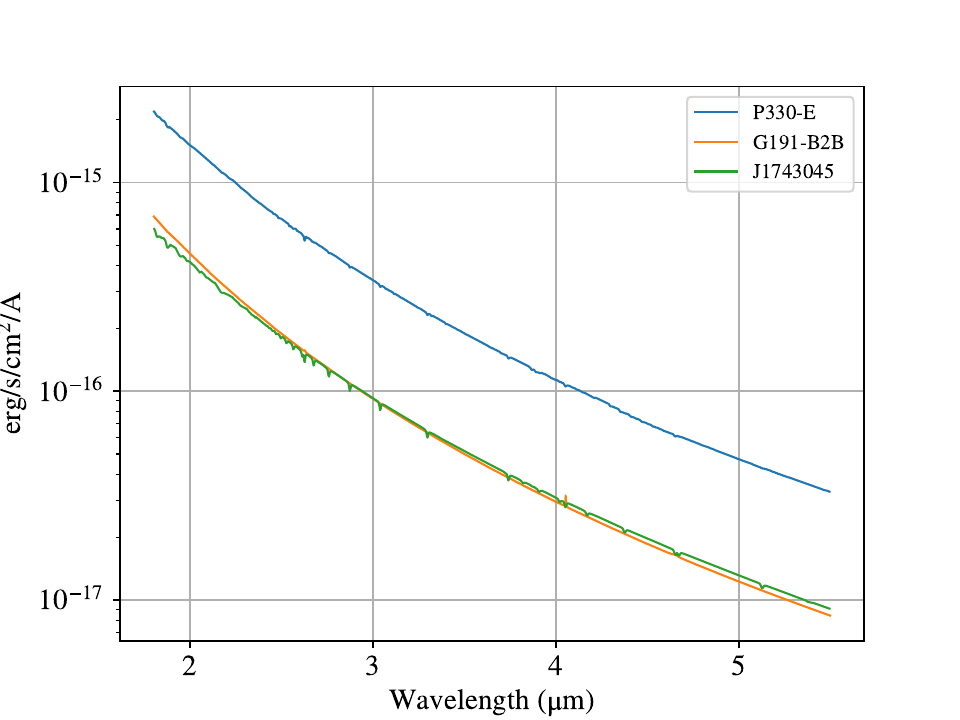}
\caption{Reference spectra used to derive the sensitivity function of the NIRCam grisms \citep[CALSPEC,][]{Bohlin14}. \label{P330_ref}}
\end{figure*}

The flux calibration of the NIRCam grism first orders was performed by observing a set of flux standards and using the trace geometry and wavelength dispersion described above to extract its spectra. These sources are G191-B2B, P330-E, and J1743045 and these are described in some details in \citet{Gordon22}. We used data obtained as part of programs 01076, 01536, 01537, 01538, 04498, and 06606 when these sources were observed using multiple combinations of Module, cross-filter, and grism.
The goal of the flux calibration is to derive the conversion from observed DN/s to physical units as a function of wavelength. The resulting sensitivity function is defined as the ratio of the 1D extracted spectrum (in DN/s) to the fiducial flux-calibration spectrum, as shown in Figure~\ref{P330_ref} for P330-E. Note that there are not sufficient numbers of observations of the +2 orders to re-derive the flux calibration of the +2 order at this time.

One-dimensional spectra of P330E were extracted as follows. The source position was first determined in the available imaging (using the methodology described in Section~\ref{sec:wave}) and then used to compute the expected trace location in the WFSS observations via Equations~\ref{forward}. Several rows (or columns, depending on the grism orientation) centered on the computed trace were summed to produce the 1D spectra. The local dispersed background, estimated across the sparse field, was subtracted from each observation. Accurate background subtraction is essential, as any over– or underestimation directly biases the measured count rate. We found that the background could be reliably estimated by computing the median signal from pixels located 50–150 pixels away on either side of the trace in the cross-dispersion direction. SBE simulations of the dispersed field were used to avoid pixels contaminated by other spectra, and we masked pixels flagged in the data quality array. This approach yielded background estimates (in DN/s) as a function of wavelength, which were then subtracted from the extracted spectra.

Of the three observed flux calibrators, G191-B2B is in a relatively crowded field which reduces it effectiveness to flux calibrate WFSS observations, and J1743045 is a little under three times the brightness of P330-E, leaving P330-E to be the primary source used here to establish the flux calibration of the NIRCAM WFSS mode. Using very narrow extractions to minimize spectra contamination from nearby sources, we note that all three sources produced identical sensitivities ($\approx 1\%$) for combinations of Module, Grism, and when using the F322W2 or the F444W cross filters. Future observations and analysis will include additional targets as reliable data can be obtained.

For each combination of cross-filter, module, and grism, we produced mean spectra of P330E using 19 different extraction half-widths, ranging from $\pm$1 to $\pm$49 pixels. These initial spectra were in units of DN/s per pixel, wavelength-calibrated using the dispersion solutions from Section~\ref{sec:wav}, and then lightly smoothed using an edge-preserving filter \citep{Koga17}.

\begin{figure*}
\center
\includegraphics[width=6.5in]{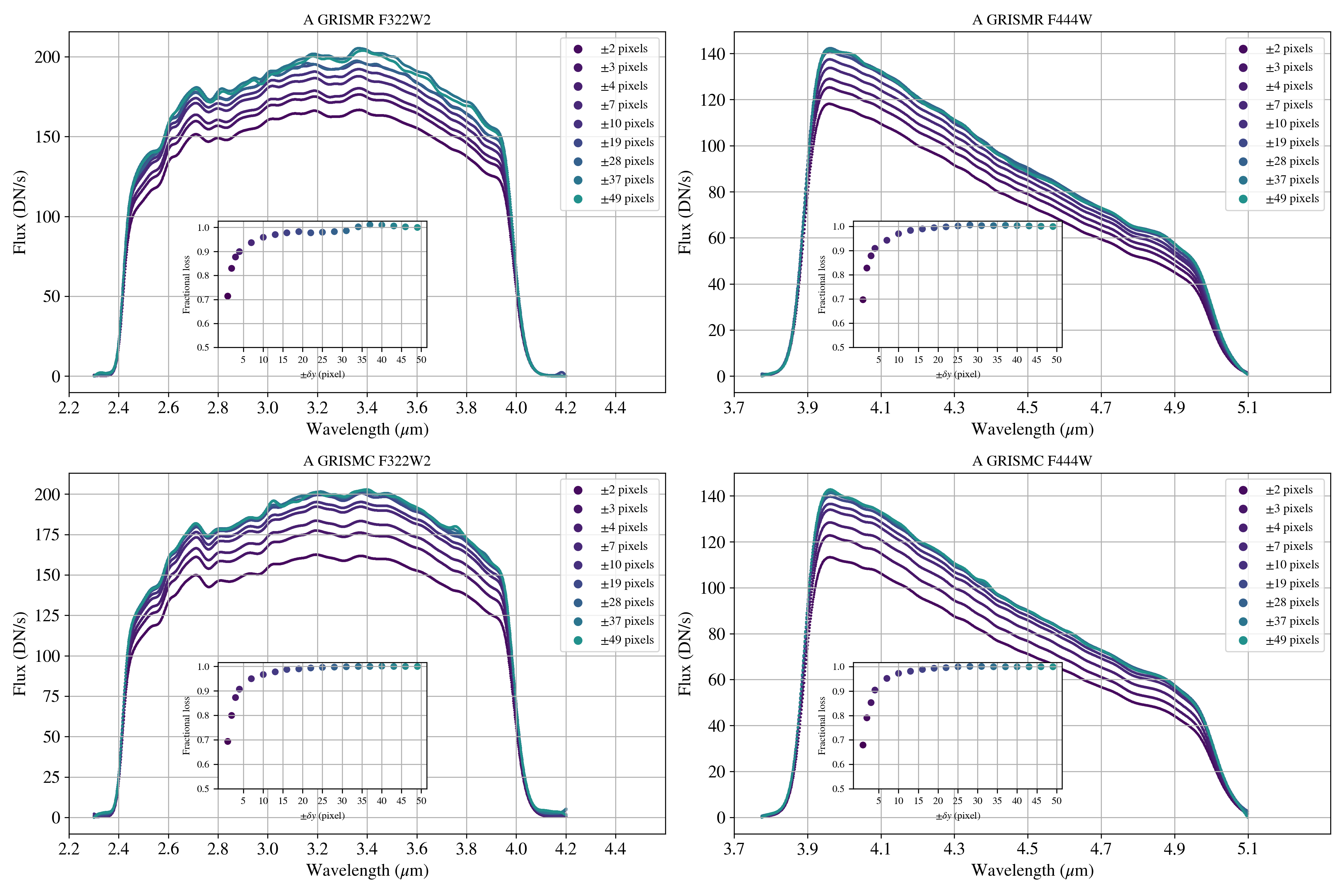}
\caption{Mean 1D spectra of P330E measured with Module~A in units of DN/s, using the F322W2 and F444W cross-filters with the GRISMR and GRISMC grisms. The legend shows the half-width extraction sizes shown. The inset plots show the average ratio of flux measured with different extraction sizes relative to the maximum extraction size of $\pm 49$ pixels. \label{aper1}}
\end{figure*}

\begin{figure*}
\center
\includegraphics[width=6.5in]{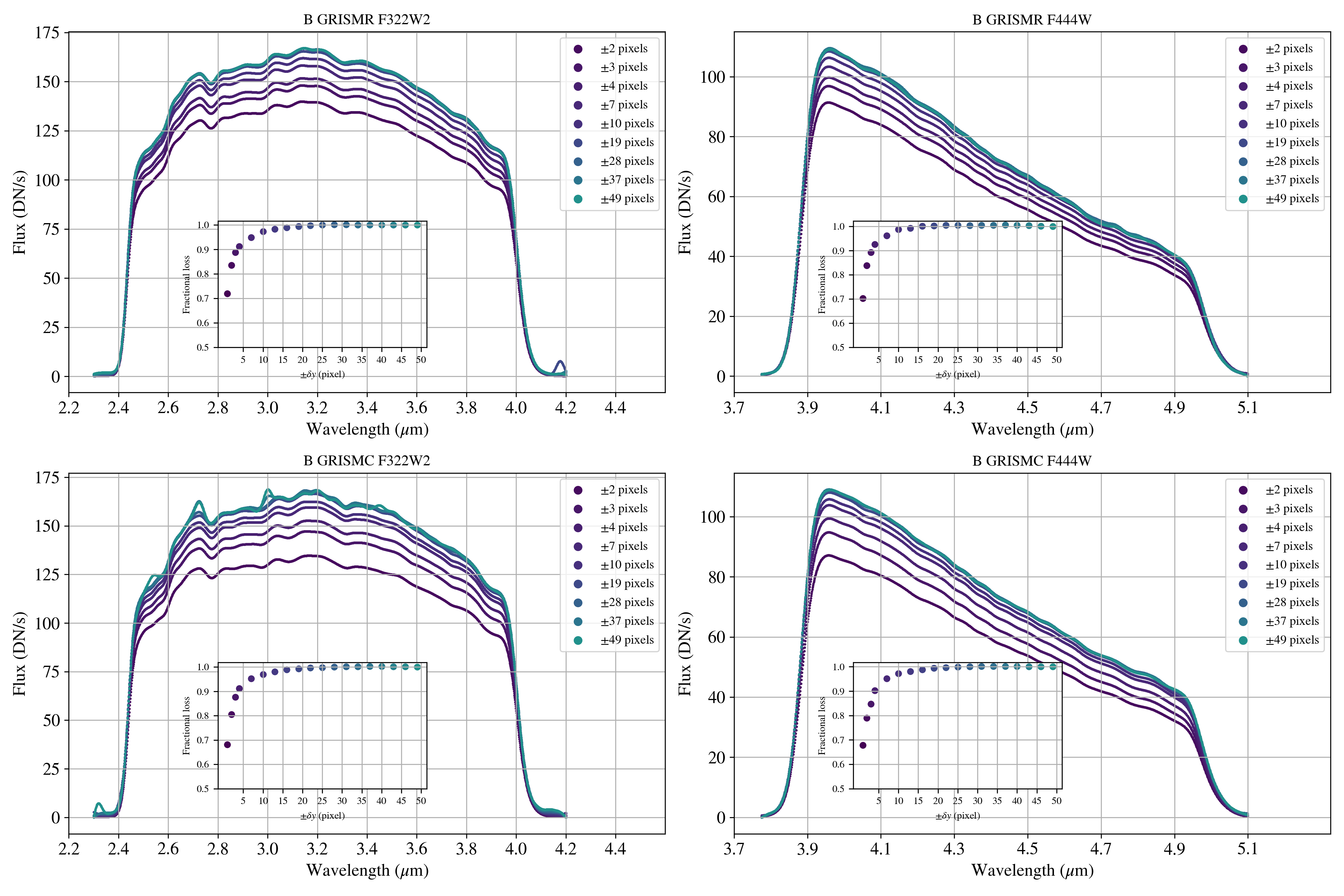}
\caption{Same as Figure~\ref{aper1}, but for Module~B. \label{aper2}}
\end{figure*}

\begin{figure*}
\center
\includegraphics[width=6.5in]{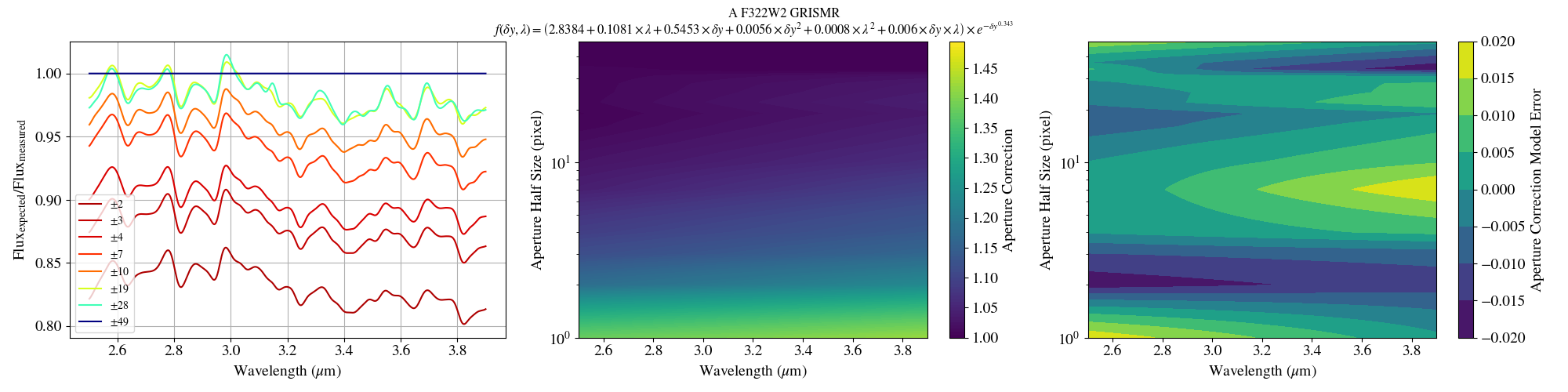}
\caption{Enclosed flux ratio for Module A, using the cross filter F322W2, GRISMR. Left Panel: Enclosed count ratio to total counts as a function of wavelength for increasing extraction aperture half-size. Middle Panel: Fitted model describing the enclosed flux ratio as a function of wavelength and aperture half-size. Right Panel: Residuals of the observed versus modeled enclosed flux ratio as a function of wavelength and aperture half-size. \label{a1}}
\end{figure*}

\begin{figure*}
\center
\includegraphics[width=6.5in]{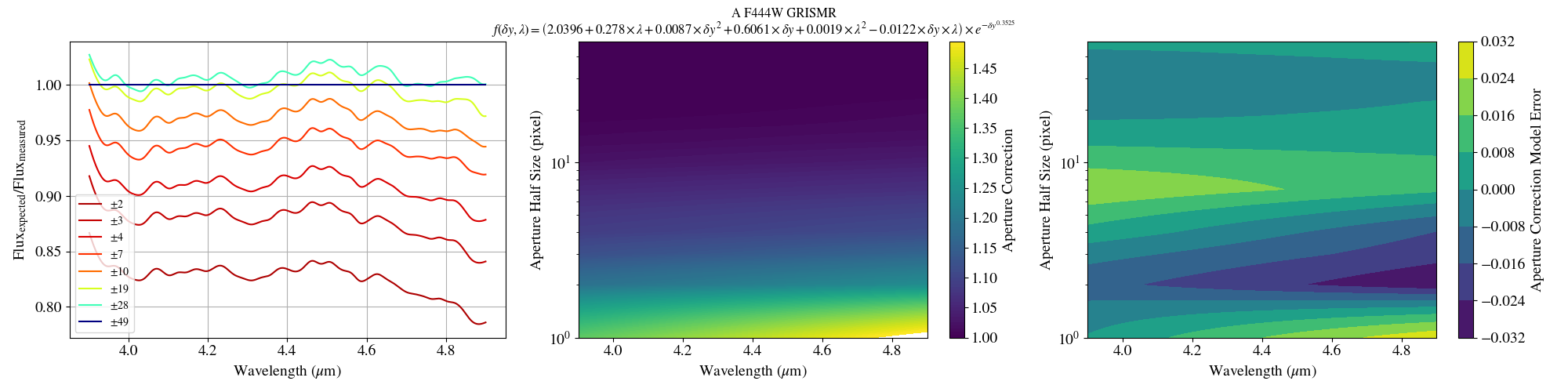}
\caption{Same as Figure~\ref{a1}, but for Module~A with cross-filter F444W and grism GRISMR. \label{a2}}
\end{figure*}

\begin{figure*}
\center
\includegraphics[width=6.5in]{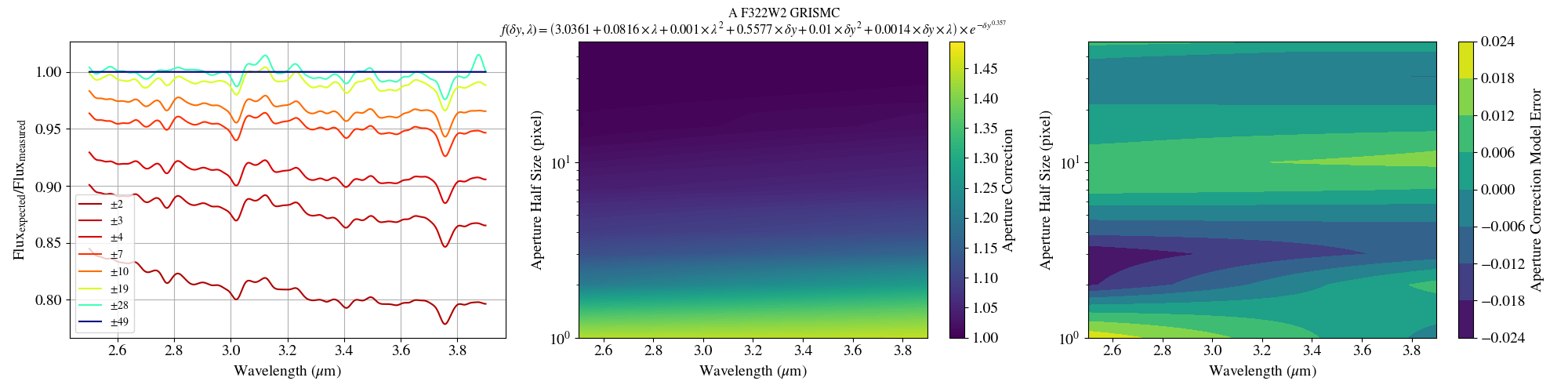}
\caption{Same as Figure~\ref{a1}, but for Module~A with cross-filter F322W2 and grism GRISMC. \label{a3}}
\end{figure*}

\begin{figure*}
\center
\includegraphics[width=6.5in]{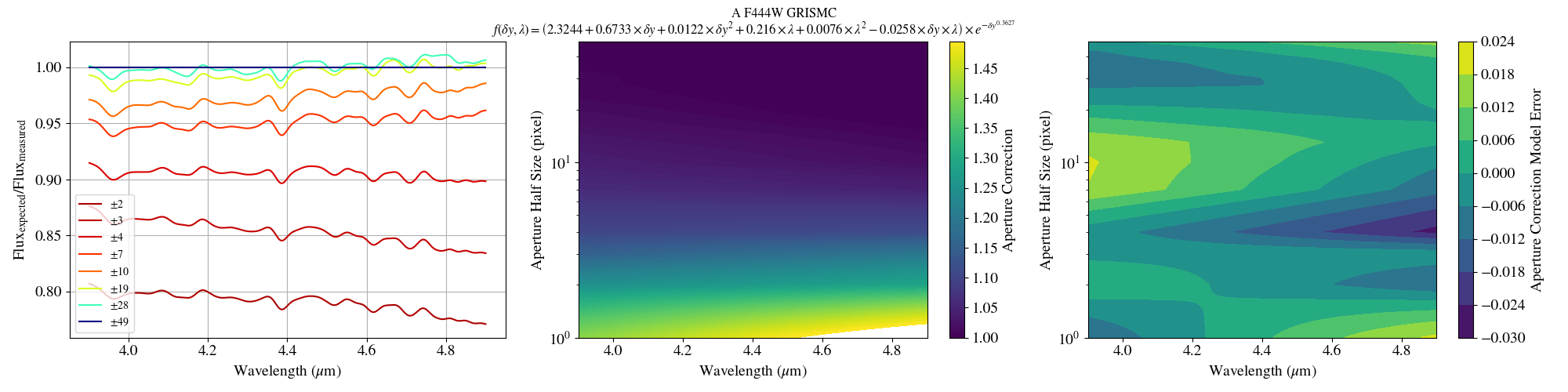}
\caption{Same as Figure~\ref{a1}, but for Module~A with cross-filter F444W and grism GRISMC. \label{a4}}
\end{figure*}

\begin{figure*}
\center
\includegraphics[width=6.5in]{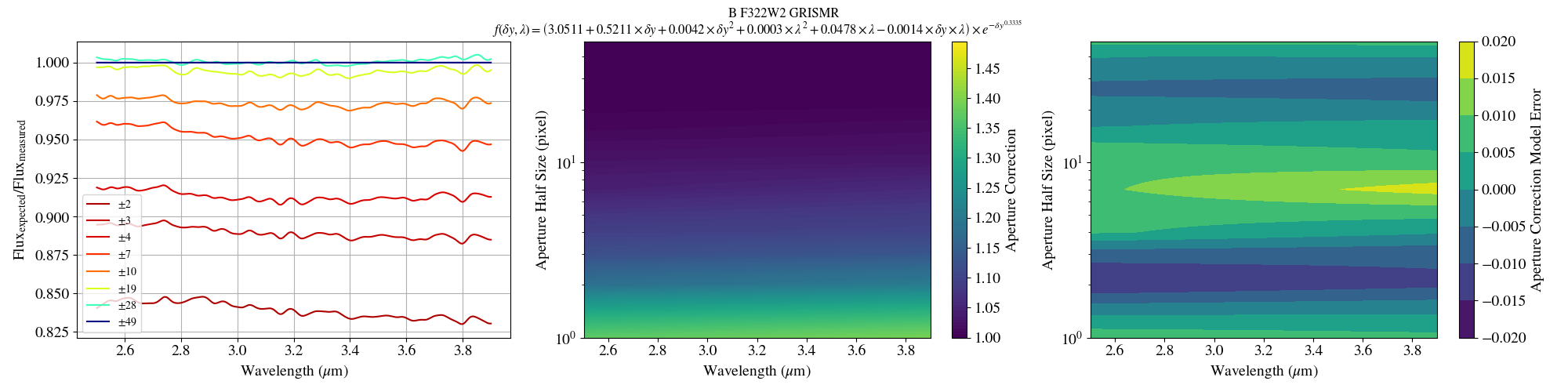}
\caption{Same as Figure~\ref{a1}, but for Module~B with cross-filter F322W2 and grism GRISMR. \label{a5}}
\end{figure*}

\begin{figure*}
\center
\includegraphics[width=6.5in]{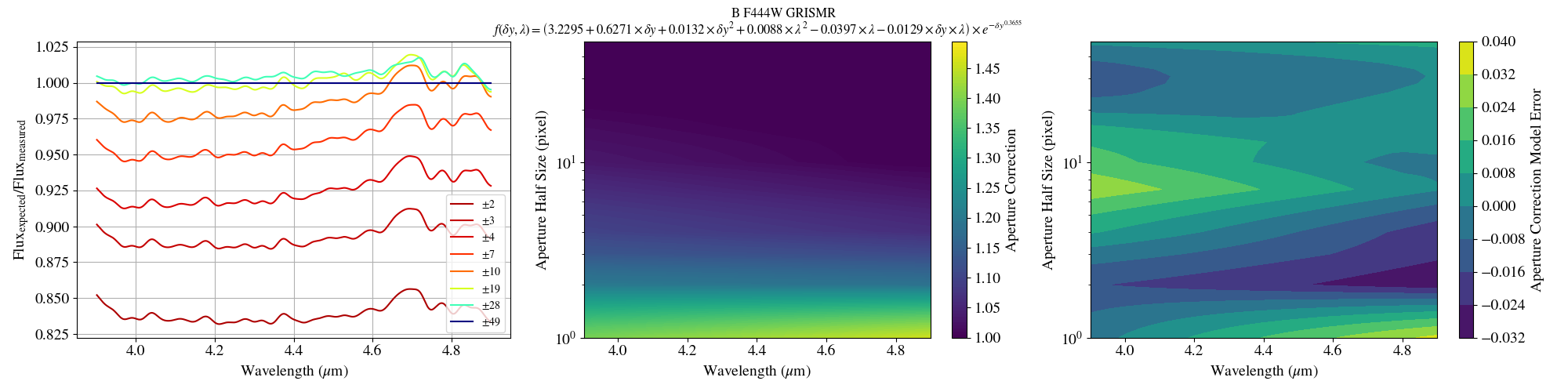}
\caption{Same as Figure~\ref{a1}, but for Module~B with cross-filter F444W and grism GRISMR. \label{a6}}
\end{figure*}

\begin{figure*}
\center
\includegraphics[width=6.5in]{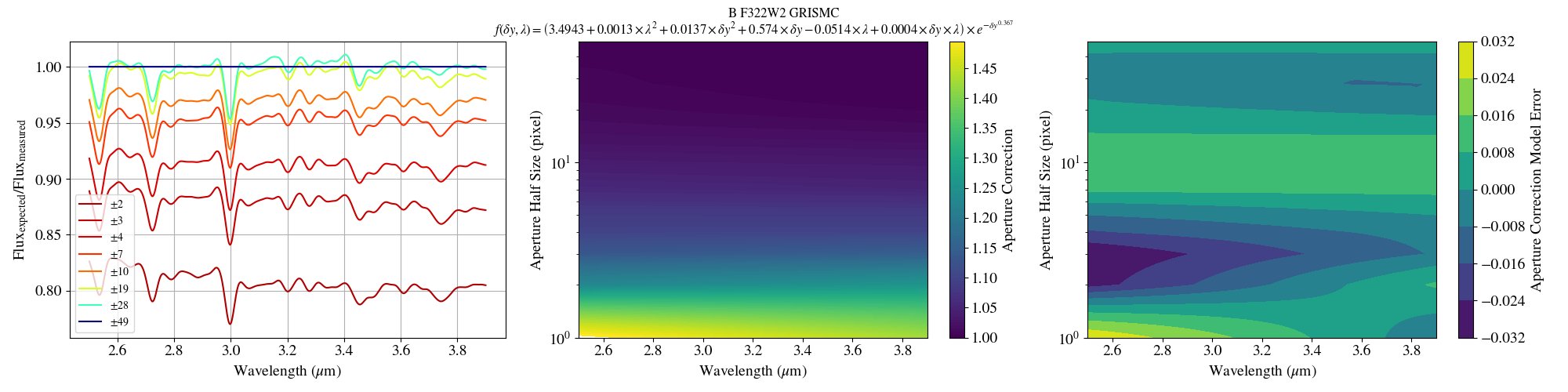}
\caption{Same as Figure~\ref{a1}, but for Module B with cross filter F322W2 and grism GRISMC. \label{a7}}
\end{figure*}

\begin{figure*}
\center
\includegraphics[width=6.5in]{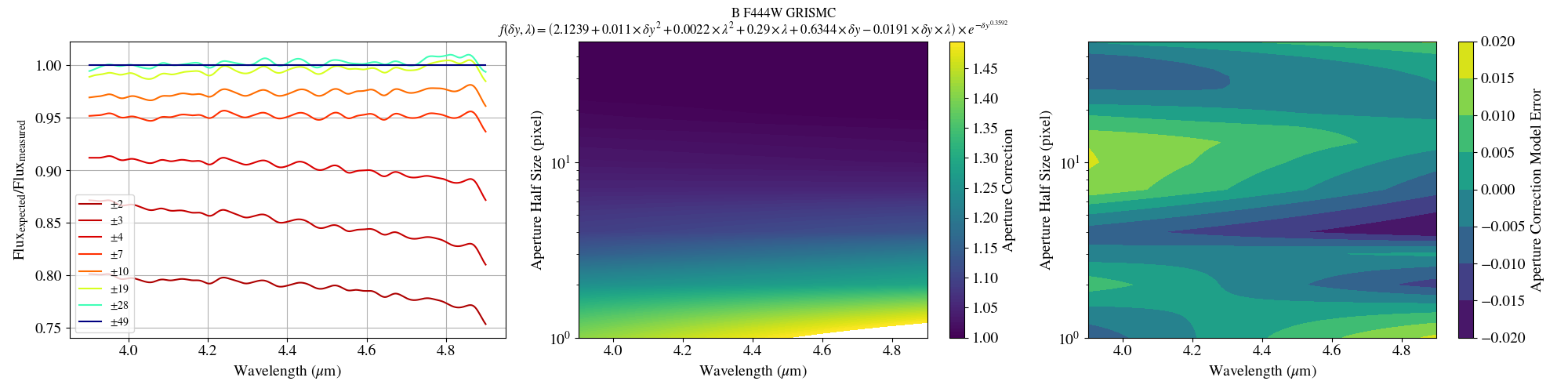}
\caption{Same as Figure~\ref{a1} but for Module B with cross filter F444W and grism GRISMC \label{a8}}
\end{figure*}

Figures~\ref{aper1} and \ref{aper2} illustrate the effect of varying the cross-dispersion extraction half-width ($\delta y$). As these figures show, a relatively wide extraction (±49 pixels) is required to capture essentially all of the flux from a point source such as P330E. For each combination of cross-filter (F322W2 or F444W), module (A or B), and grism (R or C), we extracted all usable observations—excluding those in which P330E lay too close to the detector edge to support a symmetric $\pm\delta y$ extraction. The spectra were converted from DN/s per pixel to DN/s per $\AA$\ by dividing by the pixel width in $\AA$\ (derived from the wavelength-dispersion calibration in Section~\ref{sec:wave}) and then median-combined ($\sigma \approx 1-2\%$\ for the largest extraction aperture).
To compute corrections for different extraction apertures, we modeled the change in measured flux as a function of both wavelength and extraction size. Figures~\ref{a1}–\ref{a8} show the ratio of measured flux to the maximum-aperture flux (i.e., the encircled-energy fraction) as a function of wavelength for eight extraction widths (left panels). The middle panels show the encircled-energy fraction as a function of wavelength and aperture size, along with the polynomial model used to fit the measurements. The right panels show the model residuals as a function of wavelength and extraction width. As shown in these figures, the aperture correction models agree with the observations to better than $\approx 1\%$.

\begin{rotatetable}
\begin{deluxetable}{cccccccccc}
\tabletypesize{\small}
\tablecaption{Coefficients of the Encircled Energy models for combinations of Module, cross filter, and Grism.
The function is $f(\lambda,\delta y) = e^{-{\delta y}^a_0}\times(a_1+a_2\times\delta y+a_3\times{\delta y}^2  + a_4\times\lambda + a_5\times\lambda^2+a_6\times\delta y \times\lambda)$
and returns the fraction of the total flux measured at wavelength $\lambda$\ (in $mu m$) when using an aperture size of $\delta y$ (in pixels).}
 \tablehead{\colhead{Aperture Correction Model}} 
 \tablehead{\colhead{Module} & \colhead{Filter} & \colhead{Grism} & \colhead{$a_0$} & \colhead{$a_1$} & \colhead{$a_2$} & \colhead{$a_3$}& \colhead{$a_4$} & \colhead{$a_5$} & \colhead{$a_6$}} 
 \startdata
A & F322W2 & GRISMR  & $0.34304468$ & $2.83837593$ & $0.54527674$ & $0.00562104$ & $0.10811119$ & $0.00083957$ & $0.00599417$ \\
A & F444W & GRISMR  & $0.35249737$ & $2.03959645$ & $0.60608912$ & $0.00869686$ & $0.27803718$ & $0.00190434$ & $-0.01224893$ \\
B & F322W2 & GRISMR  & $0.33346425$ & $3.05114602$ & $0.52114715$ & $0.00421798$ & $0.04781781$ & $0.00030583$ & $-0.00141107$ \\
B & F444W & GRISMR  & $0.36545683$ & $3.22950463$ & $0.62713622$ & $0.01315645$ & $-0.03970286$ & $0.00875351$ & $-0.01294893$ \\
A & F322W2 & GRISMC  & $0.3569572$ & $3.03606309$ & $0.55770594$ & $0.01000757$ & $0.08159504$ & $0.00100461$ & $0.00142145$ \\
A & F444W & GRISMC  & $0.36269208$ & $2.32436173$ & $0.67326862$ & $0.01217748$ & $0.21604084$ & $0.00763765$ & $-0.02581478$ \\
B & F322W2 & GRISMC  & $0.3670429$ & $3.49426044$ & $0.57404658$ & $0.01365254$ & $-0.05142591$ & $0.00127213$ & $0.00035836$ \\
B & F444W & GRISMC  & $0.35921864$ & $2.12389555$ & $0.63440905$ & $0.01103899$ & $0.28997333$ & $0.00220356$ & $-0.01911323$ \\
\enddata
\tablecomments{Coefficients of the Aperture Correction models\label{tab:aper}}
\end{deluxetable}
\end{rotatetable}
\newpage

\begin{figure*}
\center
\includegraphics[width=6.5in]{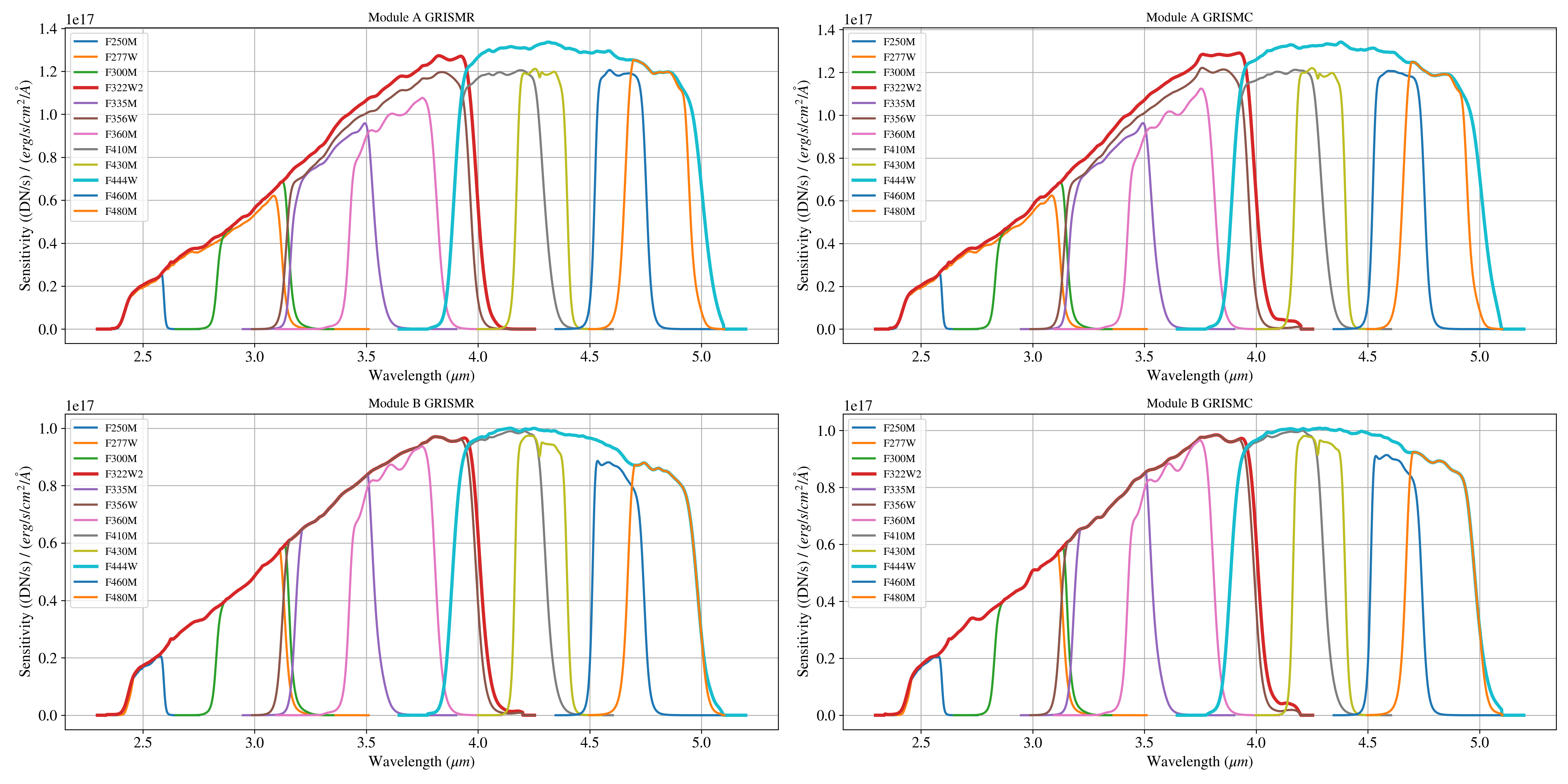}
\caption{Derived sensitivities for all combinations of cross-filters, modules, and grisms. Sensitivities measured directly from observations using F322W2 and F444W are shown with thicker lines, while those for other cross-filters are shown with thinner lines. \label{all_sens+1}}
\end{figure*}

WFSS sensitivities were derived for when using different cross-filters by dividing the median of the extracted P330E spectra, measured with a half-aperture of 49 pixels, by the reference spectrum of the calibrator shown in Figure~\ref{P330_ref}. 
The resulting +1 order sensitivities are shown in Figure~\ref{all_sens+1}.

\begin{figure*}
\center
\includegraphics[width=6.5in]{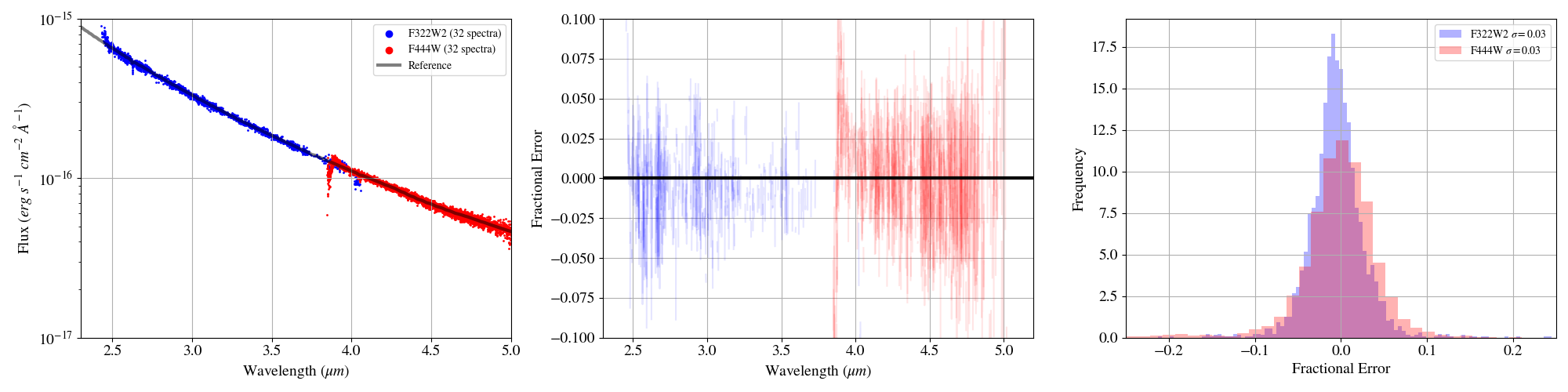}
\caption{Left Panel: Extracted and calibrated observations of P330E, taken using the F322W2 and F444W filters and the reference spectrum of P330E. F322W2 observations are shown in blue and F444W observations are shown in red. Middle Panel: The fractional deviation of our calibrated flux measurements from the reference spectrum of P330E. Right panel: histograms of the fractional deviation between calibrated observations and reference spectrum. The flux accuracy is measured to be $\sigma=3\%$ \label{P330EFinal}}
\end{figure*}

\begin{figure*}
\center
\includegraphics[width=6.5in]{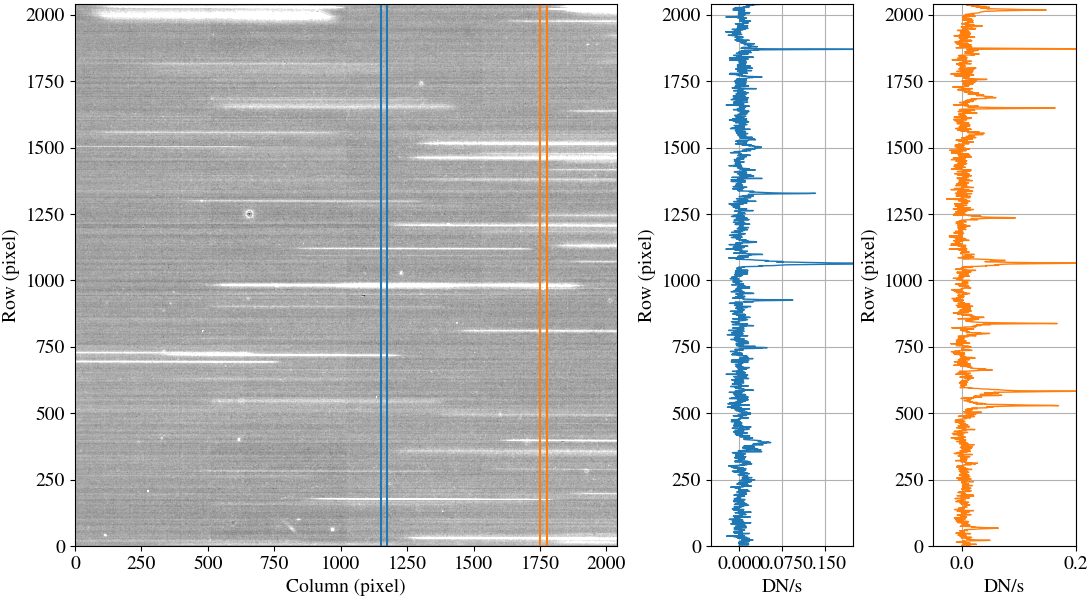}
\caption{Left Panel: A dataset from PID 01345 (CEERS), using Module A, Grism R, and F356W as a cross filter. Right Panels: Cross dispersion cuts through the data, showing emission lines. \label{example0}}
\end{figure*}

\begin{figure*}
\center
\includegraphics[width=6.5in]{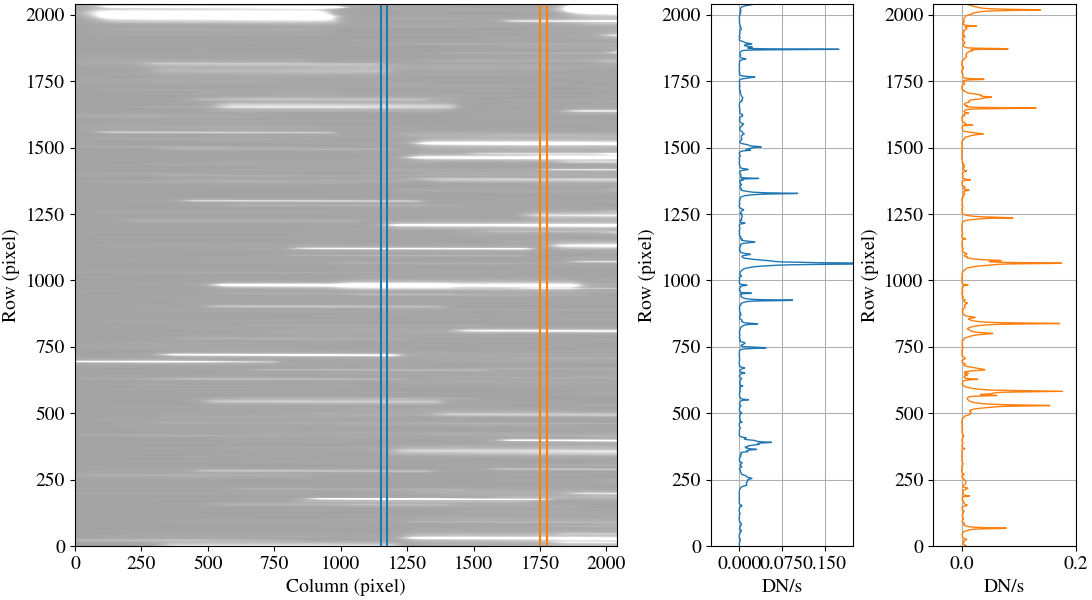}
\caption{Same as Figure \ref{example0} but now showing a simulated dataset created using the calibration described in this paper. \label{example1}}
\end{figure*}

\begin{figure*}
\center
\includegraphics[width=6.5in]{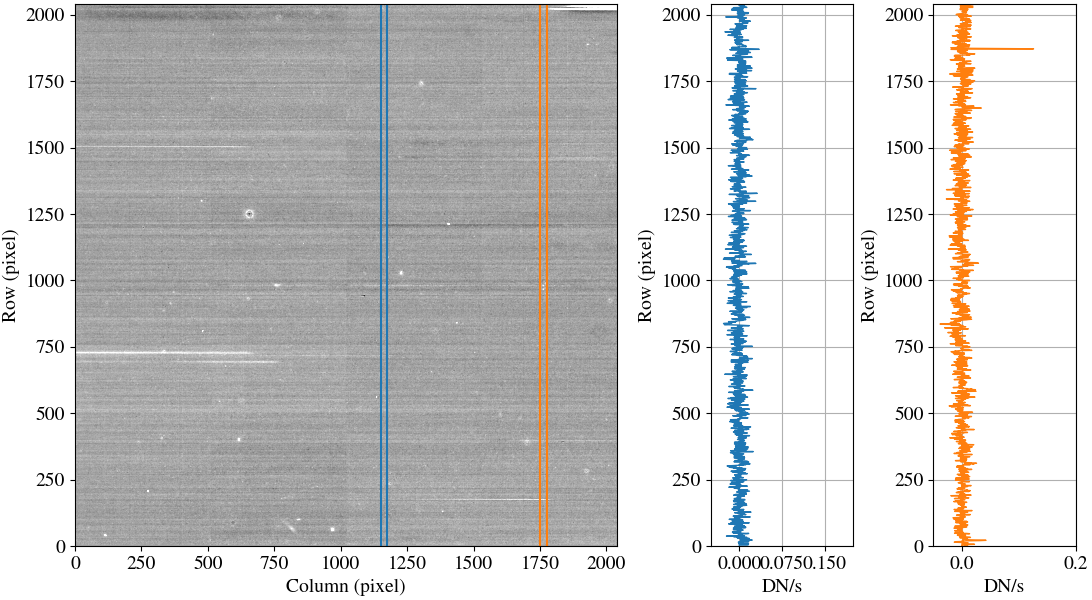}
\caption{Same as Figure \ref{example0} but showing the residuals between the observation shown in \ref{example0} and our simulated WFSS image shown in \ref{example1}. Note that the two spectra still visible at row $\approx 750$\ are two sources which were not simulated as they were not in the available direct imaging used to create the simulations shown in Figure \ref{example1}.\label{example2}}
\end{figure*}

\section{Conclusion}
We described the full process we followed to calibrate the NIRCam WFSS mode for all combinations of modules, grisms, and cross filters. Our calibration is based on the F322W2 and F444W observations.
As a final demonstration of the accuracy of the calibration presented here, we re-extracted all of the observations of P330E taken using the F322W2 and F444W cross filters (32 observations for each in total, split across Modules A and B and Grisms R and C). The data were extracted, background-subtracted (using the approach described above), and wavelength- and flux-calibrated using the calibration we derived in this paper. Figure~\ref{P330EFinal} shows a plot of all the extracted data (omitting DQ-flagged pixels and combining Module A and B as well as Grism R and C data) along with the reference spectrum of P330E. As this figure demonstrates, we achieve a $1\sigma$ flux accuracy of $3\%$, and all observations agree very well.
Figures~\ref{example0} to \ref{example2} also show an actual NIRCam WFSS observation and compare it to a simulated image produced using the WFSS calibration described in this paper, as well as the residuals between observations and simulations. An adequate calibration should allow one to produce simulated dispersed spectra that accurately reproduce observations. In particular, assuming good astrometry, one should see little to no shift between the measured positions of dispersed traces and the forward modeling of the same traces if the field dependence of the trace geometry is well calibrated. As Figure~\ref{example2} demonstrates, we obtain a clean subtraction of spectra from a real observation.
Across both Modules A and B, Grisms R and C, and cross filters F322W2 and F444W, the relative positions and variations in the shape of the NIRCam WFSS spectral traces are now calibrated to an accuracy of $0.1$ pixel over the fields of view of both modules.
We estimate the accuracy of the wavelength solution—which includes a field dependence and is based on observations of the planetary nebula SMP LMC 58 in the LMC—to be better than $1 \text{\AA}$ in all cases. Finally, the absolute flux calibration was performed using the G-type star P330E, and re-extraction of fully calibrated spectra of this source shows that the calibration described in this paper reaches an absolute flux accuracy of $3\%$.

\section*{Acknowledgements}
This work is based on observations made with the NASA/
ESA/CSA James Webb Space Telescope. The data were
obtained from the Mikulski Archive for Space Telescopes at
the Space Telescope Science Institute, which is operated by the
Association of Universities for Research in Astronomy, Inc.,
under NASA contract NAS 5-03127 for JWST.
The data described here may be obtained from the MAST archive at \dataset[doi:10.17909/55qg-gv78]{http://dx.doi.org/10.17909/55qg-gv78}.
The authors would like to thank the STScI NIRCam team and B. Hilbert in particular for their support.

\bibliography{pirzkal}{}
\bibliographystyle{aasjournalv7}



\end{document}